\documentclass{jfm}
\usepackage{graphicx}%
\usepackage{dcolumn}%
\usepackage{bm}%
\usepackage{amsmath,amssymb,setspace,graphics}%
\usepackage{rotating}%
\usepackage{latexsym}%
\usepackage[usenames]{color}%
\usepackage{natbib}

\ifCUPmtlplainloaded \else
  \checkfont{eurm10}
  \iffontfound
    \IfFileExists{upmath.sty}
      {\typeout{^^JFound AMS Euler Roman fonts on the system,
                   using the 'upmath' package.^^J}%
       \usepackage{upmath}}
      {\typeout{^^JFound AMS Euler Roman fonts on the system, but you
                   dont seem to have the}%
       \typeout{'upmath' package installed. JFM.cls can take advantage
                 of these fonts,^^Jif you use 'upmath' package.^^J}%
      }
  \else
  \fi
\fi

\ifCUPmtlplainloaded \else
  \checkfont{msam10}
  \iffontfound
    \IfFileExists{amssymb.sty}
      {\typeout{^^JFound AMS Symbol fonts on the system, using the
                'amssymb' package.^^J}%
       \usepackage{amssymb}%
         \let\leq=\leqslant
         \let\geq=\geqslant
      }{}
  \fi
\fi

\ifCUPmtlplainloaded \else
  \IfFileExists{amsbsy.sty}
    {\typeout{^^JFound the 'amsbsy' package on the system, using it.^^J}%
     \usepackage{amsbsy}}
    {\providecommand\boldsymbol[1]{\mbox{\boldmath $##1$}}}
\fi

\newsavebox{\astrutbox}
\sbox{\astrutbox}{\rule[-5pt]{0pt}{20pt}}

\newcommand\etal{\mbox{\textit{et al.}}}

\newcommand{\mathsym}[1]{{}}
\newcommand{\unicode}[1]{{}}

\title[Hydrodynamic forces on porous particles]{Hydrodynamic forces on steady and oscillating porous particles}

\author[Santtu T. T. Ollila, Tapio Ala-Nissila and Colin Denniston]%
{S\ls A\ls N\ls T\ls T\ls U\ns T.\ns T.\ns O\ls L\ls L\ls I\ls L\ls A$^{1,2}$,\ns T\ls A\ls P\ls I\ls O\ns A\ls L\ls A\ls -\ls N\ls I\ls S\ls S\ls I\ls L\ls A$^2$\break \and C\ls O\ls L\ls I\ls N\ns D\ls E\ls N\ls N\ls I\ls S\ls T\ls O\ls N$^1$\thanks{Email address for correspondence: cdennist@uwo.ca}}

\affiliation{$^1$Department of Applied Mathematics, The University of Western Ontario, London, Ontario, Canada N6A 5B8\\[\affilskip]
$^2$Department of Applied Physics, Aalto University School of Science and Technology, P.O. Box 11000, FIN-00076 Aalto, Espoo, Finland}

\date{\today}

\begin{document}

\maketitle
\begin{abstract}
We derive new analytical results for the hydrodynamic force exerted on a sinusoidally oscillating porous shell and a sphere of uniform density in the Stokes limit. The coupling between the spherical particle and the solvent is done using the Debye-Bueche-Brinkman model, {\it i.e.} by a frictional force proportional to the local velocity difference between the permeable particle and the solvent. We compare our analytical results and existing dynamic theories to Lattice-Boltzmann simulations of full Navier-Stokes equations for the oscillating porous particle. We find our analytical results to agree with simulations over a broad range of porosities and frequencies.
\end{abstract}
\begin{keywords}
\end{keywords}

\section{Introduction} 

The degree of porosity affects sedimentation and aggregation dynamics of suspended particles. For instance, high-porosity, low-density particles have been found suitable for efficient delivery of therapeutics into the systemic circulation through inhalation. Such treatment is made possible by careful engineering of porous particle structure and dynamics to circumvent pulmonary mechanisms for removing deposited particles~\citep{RLanger97}. Besides clinical applicability, models of fluid flow through porous media have been developed and tested, for instance, to improve the efficiency of oil recovery by fluid injection~\citep{Baba03} and ultrasonic waves~\citep{AA-H06}, to characterize structural properties in pulp and paper science~\citep{RGGAKKR04} and to identify conditions that cause colloid detachment from surfaces in porous media~\citep{BG00}. The response to the surrounding flow depends on the mass distribution within the porous particle. Steady-state response to hydrodynamic forces and torques is well understood, but the dynamics of permeable particles still poses several unanswered questions. Only recently has it become possible to study diffusive properties of concentrated suspensions of permeable particles~\citep{ACE-JNW10}. Porosity together with understanding of hydrodynamic forces in a corrugated nanochannel could be used to take advantage of size-dependent transport properties~\citep{Stein09} and separation~\citep{Fu07} of nanofluids~\citep{SvdBE10}. 

The first analytical results on steady-state drag on a permeable homogeneous sphere date back to \citet{DB48} and \citet{B47_1,B47_2}. Since then, numerous works \citep{F75,FD75a,FD75b,BS93,CF09} have been published on drag force and torque on spherical particles of different mass distributions and internal structure. For example, \citet{BS93} investigated a porous shell of finite thickness. Recently, \citet{CF09} solved a related problem where the shell was wrapped around a solid core. So far, studies on the subject have largely comprised theoretical calculations in the Stokes approximation of the Navier-Stokes equation.

In this work, we first compare the results of such calculations of steady-state quantities to computer simulations of the Navier-Stokes equations by the well-established Lattice-Boltzmann method (LB). We will show that our simulations give quantitative agreement with theoretical predictions without any adjustable fitting parameters at all levels of permeability for the steady-state case.

We then examine the dynamic case in which additional complications arise as the particle moves in the fluid in an oscillatory manner. \citet{LC04} used a perturbative expansion to find the hydrodynamic force on a slightly permeable sphere. They found significant differences in the fluid velocity around and in the hydrodynamic force on the particle in a frequency range from $1$ to $10\,\mathrm{MHz}$. \citet{VS09} generalized the original \citet{S51} result of the hydrodynamic force on a sinusoidally oscillating solid particle with a no-slip boundary condition. They formulated the problem such that changes in both the velocity and acceleration-dependent part of the dynamic force could be quantified as the frequency of the oscillation, the porosity of the particle or as the boundary condition on the surface of the particle was changed.  In the high-porosity limit, where the Brinkmann $\beta$ parameter tends to zero, the hydrodynamic drag force on the particle should approach zero.  However, the model of \citet{VS09} gives a finite hydrodynamic drag force on the particle for all values of $\beta$. In this work, we present a new analytical derivation for both a porous shell and a uniform density porous sphere.  Our results give a physically consistent hydrodynamic force for all values of the coupling parameter. We then find our new result to provide better overall agreement with simulations of a particle oscillating in the fluid. Our tests are performed at $0.06$--$28\,\mathrm{MHz}$ for particles of radii between $80$ and $700\,\mathrm{nm}$.

\section{Model}
\subsection{Time-dependent Stokes, Darcy and Brinkman equations}
We use the Debye-Bueche-Brinkman (DBB) model, which allows one to study generic hydrodynamic effects between a solvent and porous particles with few parameters. Typically, this model has been studied theoretically in the steady-state case.  However, we are interested in oscillating particles so will look at the time-dependent model.  The coupled porous particle-fluid system in the case of an incompressible fluid in the small Reynolds number ($Re$) limit, can be described by the time-dependent linearized Navier-Stokes equations
\begin{subequations}
\label{eq:fEOM}
\begin{eqnarray}
&&\nabla \cdot {\bf u} = 0\label{eq:fEOMa};\\
&&\rho \,\partial_t {\bf u} = \eta \nabla^2 {\bf u} - \nabla p + {\bf f},\label{eq:fEOMb}
\end{eqnarray}
\end{subequations}
where $\rho$ is the fluid mass density, ${\bf u}$ the fluid velocity, $\eta$ the shear viscosity and $p$ is the pressure.  Here, the presence of the porous particle is characterized by the force density,
\begin{subequations}
\label{eq:forcedensity}
\begin{eqnarray}
{\bf f} &=&\gamma n({\bf r})({\bf v}-{\bf u})\label{eq:forcedensitya};\\
n({\bf r}) &=& \left \{
\begin{array}{ll}
 \lambda, & {\bf r} \in B(t),\\
 0, & {\bf r} \notin B(t),
 \end{array}\right.\label{eq:forcedensityb}
 \end{eqnarray}
\end{subequations}
where the coupling constant $\gamma$ has units mass per time, ${\bf v}$ is the local velocity of the particle at the point ${\bf r}$, which contains contributions from centre of mass and rotational motion. The ``node'' density $n({\bf r})$, which has units of inverse volume, has a constant value $\lambda$ inside the particle and zero outside the volume $B(t)$ of the particle. Outside $B(t)$, where ${\bf f}=0$, (\ref{eq:fEOM}) is commonly referred to as the unsteady Stokes equation (it is missing the nonlinear $\rho {\bf u}\cdot\nabla {\bf u}$ term present in the full Navier-Stokes equations).

Inside $B(t)$, the fluid flow interacts directly with the nodes and (\ref{eq:fEOM}) is referred to as the DBB equation~\citep{B47_1,B47_2,DB48}. The shape of the particle $B(t)$ can be varied to give a shell, uniform-density sphere or other distribution. The product $\gamma \lambda$ is equal to $\eta \kappa^2$~\citep{ACE-JNW10}, where $\kappa^{-1}$ is the hydrodynamic screening length and $\kappa^{-2}=\eta/(\gamma\lambda)$ is the constant permeability of the particle.

The DBB equation is a mean-field description of fluid flow in the porous particle under the assumption that the particle radius $R$ is large enough compared to the mean pore size $\kappa^{-1} \sim \sqrt{\eta/(\gamma \lambda)}$. If one further neglects the Laplacian term $\eta \nabla^2 {\bf u}$ inside $B(t)$, then one arrives at the Darcy model, for which the fluid velocity inside the particle is independent of $r=\vert {\bf r} \vert$. The viscosity is assumed to be $\eta$ both inside and outside of $B(t)$.

In the frame of reference with the origin at the centre of mass of our porous particle, we define a spherical coordinate system $(r,\varphi,\theta)$ via $(x,y,z) = r(\sin\theta\cos\varphi,\sin\theta\sin\varphi,\cos\theta)$. As we only consider axisymmetric flows, the solution to (\ref{eq:fEOM}) is independent of $\varphi$, and ${\bf u}$, $p$ and the fluid stress tensor $\boldsymbol{\sigma}$ will only depend on $(r,\theta)$~\citep{G07}. Once the dependence of $\boldsymbol{\sigma} = \boldsymbol{\sigma}(r,\theta)$ is known, one may proceed to calculate the hydrodynamic force ${\bf F}$ and torque ${\bf T}$ acting at the centre of mass of our porous particle from the stress as~\citep{LandauLifschitz87}
\begin{eqnarray}
{\bf F} & = & \int_{\partial B(t)} \boldsymbol{\sigma} \cdot  \hat{{\bf e}}_r\,\mathrm{d}S;\label{eq:hForce}\\
{\bf T} & = & \int_{\partial B(t)} r \,\hat{{\bf e}}_r \times \boldsymbol{\sigma} \cdot  \hat{{\bf e}}_r\,\mathrm{d}S,\label{eq:hTorque}
\end{eqnarray}where $(\hat{{\bf e}}_r,\hat{{\bf e}}_\varphi,\hat{{\bf e}}_\theta)$ are the unit basis vectors in spherical coordinates and $\partial B(t)$ is the boundary of $B(t)$.

\begin{figure}
\centerline{\includegraphics[width=0.48\textwidth]{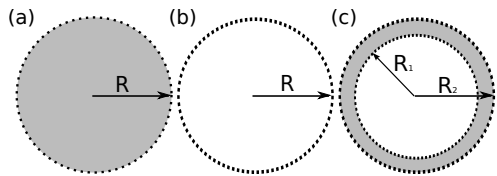}}
\caption{Schematic of the (a) uniform-density sphere with node density $n_\mathrm{Sp}({\bf r})$, (b) infinitely thin shell $n_\mathrm{Sh}({\bf r})$ and (c) a shell of finite thickness/annulus $n_\mathrm{An}({\bf r})$ which is hollow on the inside where the Stokes approximation applies.}
\label{fig:schematic}
\end{figure}

The porosity-dependent force and torque exerted on the particle by the fluid can also be calculated directly using Newton's third law from the integral of the negative of the force density on the fluid:
\begin{eqnarray}
{\bf F} &=& \int_{B(t)} -{\bf f}\,\mathrm{d}^3x = \int_{B(t)}-\gamma\, n({\bf r})({\bf v}-{{\bf u}})\,\mathrm{d}^3 x\nonumber\\
&=& - \int_{B(t)}\!\!\gamma\, n({\bf r})\Bigr(\!{\bf v}_\mathrm{cm} + {\bf w} \! \times \! ({\bf r} - {\bf r}_\textrm{cm}) \!-\! {\bf u}({\bf r})\!\Bigl)\mathrm{d}^3 x.\label{eq:F}
\end{eqnarray}
where ${\bf w}$ is the angular velocity of the particle and ${\bf v}_\mathrm{cm}$ its centre-of-mass velocity. Schematics of the spherical particles we consider are shown in figure~\ref{fig:schematic}. The hydrodynamic torque ${\bf T}$ on the particle then reads
\begin{eqnarray}
{\bf T} & =& \int_{B(t)} ({\bf r} - {\bf r}_\mathrm{cm}) \times (-{\bf f})\,\mathrm{d}^3 x\nonumber\\
& =& -\int_{B(t)} \gamma\, n({\bf r}) ({\bf r} - {\bf r}_\mathrm{cm})
\Bigr({\bf v}_\mathrm{cm} + {\bf w} \times ({\bf r} - {\bf r}_\textrm{cm}) - {\bf u}({\bf r})\Bigl).\label{eq:T}
\end{eqnarray}
In this work, the particle is either held fixed, ${\bf v}_\mathrm{cm} = 0$, or its velocity is set to be sinusoidal along the $z$ axis. We separate the analytical solution to (\ref{eq:fEOM}) to parts inside and outside $B(t)$ following conventions of~\citet{F75}. The parts are matched as detailed in sections \ref{sec:exactshell} and \ref{sec:exactsphere} by requiring the velocity and stress fields to be equal on the boundary of $B(t)$.

\subsection{Previous work}
A few other authors have also examined the time-dependent case.  \citet{LC04} performed a perturbative expansion to find the force exerted on a rigid, weakly permeable sphere of radius $R$ oscillating in an incompressible fluid. The dimensionless perturbation parameter $\epsilon = (\kappa R)^{-1}$ was studied in the range $\left[0,0.05\right]$, and they consistently neglected terms of order $\mathcal{O}(\epsilon^2)$ and higher. They modeled the porous sphere by applying the Beavers-Joseph-Saffman (BJS) boundary condition on its surface to order $\epsilon$. They solved the homogenized unsteady Stokes equation by assuming the particle to be impermeable enough so that the flow external to $B(t)$ cannot penetrate the particle interface.

More recently, \citet{VS09} also studied an oscillating sphere of uniform permeability. They used (\ref{eq:fEOM}) outside the sphere (as ${\bf f}({\bf r}) \equiv 0$ for ${\bf r}\,\notin\, B(t)$) and inside the sphere, their dynamic equation,
\begin{equation}
\label{eq:VSinside}
\rho \partial_t \tilde{{\bf u}} = \eta \nabla^2 \tilde{{\bf u}} - \nabla p - \eta \kappa^2 \tilde{{\bf u}},
\end{equation}did not contain the particle velocity. This equation corresponds to (\ref{eq:fEOM}) and (\ref{eq:forcedensity}) inside $B(t)$ only if ${\bf v}\equiv 0$ in (\ref{eq:forcedensity}). Instead of introducing the particle-fluid interaction in (\ref{eq:VSinside}), they introduced it as a boundary condition
\begin{equation}
\label{eq:VSBC}
{\bf u}({\bf r}= R\hat{{\bf e}}_r) - {\bf v} = \tilde{{\bf u}}({\bf r}= R\hat{{\bf e}}_r),
\end{equation}where the particle velocity is ${\bf v}=v_0 e^{i \omega t} \hat{{\bf e}}_z$, $\tilde{{\bf u}}$ refers to the fluid velocity inside $B(t)$ and ${\bf u}$ to that outside $B(t)$. They also required continuity of components of the stress tensor $\sigma_{rr}$ and $\sigma_{r\theta}$ at ${\bf r}=R\hat{{\bf e}}_r$ when the Laplacian term was included in (\ref{eq:VSinside}) and continuity of pressure when it was neglected. Note that boundary condition (\ref{eq:VSBC}) is {\it not} equivalent to the DBB model.  Any attempt to remove the particle velocity in the equations of motion of the DBB model (which enters in the ${\bf f}$ in (\ref{eq:fEOM})) would end up introducing the particle inertia in (\ref{eq:fEOM}) (arising from the equation of motion for the particle).  Vainshtein and Shapiro's model has neither ${\bf v}$ nor $d{\bf v}/dt$ in (\ref{eq:VSinside}).  We find it difficult to motivate Vainshtein and Shapiro's model physically and will demonstrate that (\ref{eq:VSinside}) and (\ref{eq:VSBC}) actually give a physically incorrect limit for the time-dependent case for small $\beta$.

\subsection{Simulation method}
\label{sims}
We will compare the Stokes theory results to simulations of the full Navier-Stokes equations through the well-established LB method.  
The mass and momentum conservation in a fluid are expressed at the Navier-Stokes level as~\citep{Batchelor67,LandauLifschitz87}
\begin{equation}
\partial_t \rho + \partial_{\alpha}(\rho u_{\alpha}) = 0
\label{eq:continuity}
\end{equation}and
\begin{eqnarray}
&&\partial_t(\rho u_\alpha) + \partial_\beta(\rho u_\alpha u_\beta) =  -\partial_\alpha p + f_\alpha\label{e:NavierStokes}\\
 &&\,+\partial_\beta\left(\eta\Bigl(\partial_\alpha u_\beta+\partial_\beta u_\alpha - \frac{2}{3}\partial_\gamma u_\gamma\delta_{\alpha\beta}\Bigr) +  \zeta \partial_\gamma u_\gamma\delta_{\alpha\beta}\right),\nonumber
\end{eqnarray}where $\eta$ and $\zeta$ are the shear and bulk viscosities and $p$ is the fluid pressure. In this work we will use a pressure with linear dependence on density, {\it i.e.} $p = \rho v^2_s \delta_{\alpha\beta}$, where $v_s$ is the speed of sound.  This can be viewed as an ideal gas equation of state or the first term in a Taylor expansion of the pressure about fixed density in which case $v^2_s$ is the isentropic compressibility~\citep{Kell70}. The coupling to the particle phase appears through the force density $f_\alpha$.

Our lattice Boltzmann (LB) fluid algorithm, summarized in Appendix~\ref{appendixLB}, reproduces (\ref{eq:continuity}) and~(\ref{e:NavierStokes}) in the form typical to most LB algorithms~\citep{CD98}. The shear viscosity in the model is $\eta = \rho \tau v_c^2/3$, where $v_c=\Delta x/\Delta t$ is a lattice velocity, and $\zeta = \eta(5/3-3v_s^2/v_c^2)$ \citep{SOY95}. In this paper, $\tau$ will be chosen in all cases so that $\eta = 0.02\,\mathrm{g}\,\mathrm{cm}^{-1}\mathrm{s}^{-1}$, twice the viscosity of water, and $\rho = 1\,\mathrm{g}\,\mathrm{cm}^{-3}$. The speed of sound $v_s$ is chosen to be $v^2_s=v_c^2/3$ ($v_s<v_c$ is required for stability in LB algorithms). This is sufficiently large so that the fluid is approximately incompressible for steady-state situations (largest variation in $\rho <0.1\%$).  In this work, unless stated we use a time step $\Delta t = 1\,\textrm{ns}$ and a mesh resolution of $\Delta x = 100\,\mathrm{nm}$.

In the simulation, the node density is discrete, 
\begin{equation}
n({\bf r}) = \sum_{i=1}^N \delta({\bf r} - {\bf r}_i).
\end{equation}
That is, the particle is an extended spherical object consisting of $N$ nodes/constituents positioned at ${\bf r}_i$ and moving at velocity ${\bf v}_i$ around the centre-of-mass coordinate ${\bf r}_\textrm{cm}$ whose velocity is ${\bf v}_\mathrm{cm}$.  The nodes are coupled to the fluid lattice locally by weighted interpolation, which has been used for polymers consisting of point particles~\citep{AD98}, a nanowire immersed in a nematic liquid crystal~\citep{SD07} and most recently for polymers consisting of composite shells~\citep{ODKA11} like those in this work. The method is similar to Peskin's immersed boundary method~\citep{P02}.  The LB method has been used successfully to model fluid flow in porous media~\citep{KZC02}.

We generate the node distribution for the shell ($n_\mathrm{Sh}({\bf r})$) in simulation by using the atomic coordinates of spherical fullerenes, consisting of $N$ carbons, scaled so the nodes sit at radius $R$. The number of nodes is chosen sufficiently large for a given $R$ to guarantee a node placement denser than the resolution of the underlying lattice. This is necessary in order to resolve the spherical shape as the value of the coupling constant $\gamma$ is increased away from the free-draining limit. The uniform-density sphere (\ref{eq:nSphere}) and the shell of finite thickness (\ref{eq:nAnnulus}) discussed below are generated by placing nodes at intervals $\Delta x_n$ on a cubic lattice at coordinates ${\bf r}$ which fulfill $0 \leq \vert {\bf r} - {\bf r}_\mathrm{cm} \vert \leq R$ and $R_1 \leq \vert {\bf r} - {\bf r}_\mathrm{cm} \vert \leq R_2$, respectively.

\section{Analytical Results}
Next, we summarize and present calculations of the components of ${\bf F}$ and ${\bf T}$ in typical steady states for particles with different node densities $n({\bf r})$ that we will later compare to our simulation result. We then look closely at the dynamic problem of the particle oscillating sinusoidally in the fluid.
\subsection{Steady-state solutions}
In this subsection of the paper, we summarize previously derived steady-state solutions to (\ref{eq:fEOM}), where $\partial_t {\bf u}$ is assumed to be zero, for spherical particles of different node densities that will act as limiting cases of the oscillating particle solution. The particle is assumed to be fixed in place and the far-field velocity constant.

The simplest experimentally relevant case is that of a {\it sphere of uniform node density} ($\Theta$ is the Heaviside step function),
\begin{equation}
n_\textrm{Sp}({\bf r}) = \lambda\Theta(R-r) = N\Bigl(\frac{4}{3}\pi R^3\Bigr)^{-1}\Theta(R-r),
\label{eq:nSphere}
\end{equation}
immersed in a background fluid whose far-field velocity ${\bf u}_\infty = {\bf u}(r\rightarrow\infty) = U_0 \hat{{\bf e}}_z$. As mentioned in Section~\ref{sims}, $N$ is large enough in the simulations so that the discrete nodes are spaced closer together than the fluid mesh spacing to approximate a uniform density reasonably. \citet{DB48} solved the problem, in the context of the uniform density sphere being a model for a polymer in solution, for arbitrary values of the dimensionless parameter $\beta = \kappa R = \sqrt{\gamma \lambda/\eta}R$. The product $\gamma \lambda$ describes the strength of coupling between the phases. In simulation, one may lower the node density $\lambda$ and increase the coupling parameter $\gamma$ while keeping the product the same without significantly changing the results as long as the node placement is sufficiently dense to resolve the shape of the object~\citep{OSD11}. Debye and Bueche presented their original solutions as functions of $\beta$ as defined here without any reference to the size of the node itself, but only to $R$ and a so-called shielding length $\sqrt{\eta / (\lambda\gamma)}$, which is also referred to as mean pore size by \citet{ACE-JNW10}. Debye and Bueche also related the shielding length to the slip length in the Navier slip boundary condition~\citep{DB48}. Moreover, Bueche found the drag force on the spherical uniform-density ''polymer" to be ${\bf F} = F\hat{{\bf e}}_z$, where
\begin{eqnarray}
\frac{F}{6\,\pi\,\eta R\, U_0} & = & \frac{2 \beta^2 G_0(\beta)}{2\beta^2 + 3 G_0(\beta)};\label{eq:ssSphereF}\\
G_0(\beta) & \equiv & 1 - (1/\beta)\tanh(\beta)\label{eq:reductionfactor}.
\end{eqnarray}Both the limit of zero, $\beta\rightarrow 0$, and infinite, $\beta\rightarrow\infty$, coupling in (\ref{eq:ssSphereF}) give the intuitive results $F\rightarrow 0$ and the Stokes formula $F\rightarrow F_S \equiv 6\,\pi\,\eta\,R\,U_0$ first derived by Stokes for an impermeable sphere with a no-slip boundary condition on its surface~\citep{S80}. At low porosity (large $\beta$), (\ref{eq:ssSphereF}) can be approximated by
\begin{align}
\frac{F}{F_S} & \approx \frac{2\beta^2}{2\beta^2 + 3},\label{eq:ssSphereFapprox}
\end{align}which differs from (\ref{eq:ssSphereF}) by less than $10\%$ for $\beta > 10.9$. This approximation will be relevant for what follows so we treat it here in more detail. \citet{ST70} arrived at (\ref{eq:ssSphereFapprox}) directly by assuming Darcy's law (see text below (\ref{eq:forcedensity})) to apply in the form
\begin{align}
\label{eq:Darcy}
{\bf u}({\bf r} = R\hat{{\bf e}}_r) & = g U_0 \hat{{\bf e}}_z = -\eta \kappa^2 \nabla p({\bf r} = R\hat{{\bf e}}_r)
\end{align}on the surface $\partial B(t)$. By matching the pressure based on (\ref{eq:Darcy}) and that based on the stream function for a solid sphere (see {\it e.g.} \citet{LandauLifschitz87}), they found $1-g=2\beta^2/(2\beta^2+3)$ and hence they called $g$ the permeation coefficient. This shows that the function $G_0(\beta)$ appears due to the Laplacian term of (\ref{eq:fEOMb}) included in the Brinkman model but not in the Darcy model, which is also apparent from the explicit calculation of \citet{F75}.

Alternatively, one may place the uniform-density sphere in a flow of constant shear rate $Q$ and constrain it not to rotate in which case the particle is subjected to drag torque ${\bf T} = T {\bf e}_n$, where ${\bf e}_n$ is a normal vector perpendicular to the shear plane. Its magnitude $T=\vert {\bf T} \vert$ is given by~\citep{FD75a}
\begin{align}
\frac{T}{4\,\pi\,\eta\,R^3 Q} = 1+ \frac{3}{\beta^2} - \frac{3\coth \beta}{\beta}\label{eq:ssSphereT}.
\end{align}We define the Stokes torque~\citep{Goldman67b} $T_S \equiv 4\,\pi\,\eta\,R^3 Q$ and find $T \rightarrow 0$ as $\beta \rightarrow 0$ and $T\rightarrow T_S$ as $\beta \rightarrow \infty$. Equation~(\ref{eq:ssSphereT}) is found as the solution to the full Brinkman problem from the  mean-field theory of \citet{FD75a} or as the limit of vanishing hard core for a coated particle~\citep{CF09}.

Felderhof \& Deutch have written a series of publications~\citep{FD75a,FD75b,F75} on frictional properties of dilute polymer solutions in which they show how the macroscopic \citet{DB48} results for the hydrodynamic friction coefficients are obtained as mean-field approximations from a microscopic theory by \citet{KR48}. The term mean-field is used here as the average flow velocity ${\bf u}({\bf r})$, average pressure $p({\bf r})$ and average force density ${\bf f}({\bf r})$ are taken over the statistical distribution $P({\bf r}_1,\ldots,{\bf r}_N)$ of the node positions, which were considered as segments making up the polymer~\citep{FD75a}. Felderhof and Deutch's work is significant in that they considered more general density distribution than $n_\mathrm{Sp}({\bf r})$ in (\ref{eq:nSphere}).

In particular, they considered~\citep{FD75b,F75} {\it an infinitesimally thin shell} for which the node density is given by
\begin{equation}
\label{eq:nShell}
n_\textrm{Sh}({\bf r}) = \lambda_\mathrm{Sh} \delta(r-R) \equiv N (4\,\pi R^2)^{-1} \delta(r-R),
\end{equation}where $\lambda_\mathrm{Sh}$ is the uniform surface density. Such shells are of particular interest in biophysical problems such as leaky vesicles and encapsulated drug delivery.  They calculated the shell to experience a drag force and torque equal to
\begin{subequations}
\label{eq:ssShell}
\begin{align}
\frac{F}{F_S} & = \frac{2\beta^2}{2\beta^2 + 9}\label{eq:ssShellF};\\
\frac{T}{T_S} & = \frac{\beta^2}{\beta^2 + 9}\label{eq:ssShellT}
\end{align}
\end{subequations}in the same setting that gave (\ref{eq:ssSphereF}) and (\ref{eq:ssSphereT}) for the uniform-density sphere. We emphasize that $\beta$ in (\ref{eq:ssShell}) is still equal to $R\sqrt{\gamma \lambda/\eta}$, where $\lambda$ from (\ref{eq:nSphere}) guarantees correct units. The small difference between (\ref{eq:ssSphereFapprox}) and (\ref{eq:ssShellF}) suggests that a low-porosity shell and sphere (large $\beta$) might be difficult to distinguish from one another based on drag force.

We have derived both results of (\ref{eq:ssShell}) (they are also limiting cases of the dynamic calculation we describe in the next subsection) for a stationary particle by requiring the force and torque based on the coupling and the fluid stress to match locally at every point on the shell,
\begin{subequations}
\label{eq:localF}
\begin{align}
-\gamma \lambda_\mathrm{Sh}(-{\bf u}) & = \boldsymbol{\sigma} \cdot \hat{{\bf e}}_r;\label{eq:localF:a}\\
- R\hat{{\bf e}}_r \times \gamma \lambda_\mathrm{Sh}(-{\bf u}) & = R\hat{{\bf e}}_r \times \boldsymbol{\sigma} \cdot \hat{{\bf e}}_r\label{eq:localF:b}.
\end{align}
\end{subequations}The fluid velocity field ${\bf u}$ and stress $\boldsymbol{\sigma}$ that go into (\ref{eq:localF}) are based on known stream functions~\citep{Batchelor67,LandauLifschitz87} whose constants are left arbitrary to be determined by imposing (\ref{eq:localF}).

\citet{BS93} studied {\it a shell of finite thickness} for which the node density can be written as ($R_2 > R_1$)
\begin{align}
n_\textrm{An}({\bf r}) & = \lambda_\mathrm{An} \Bigl(\Theta(R_2-r)-\Theta(R_1-r)\Bigr);\label{eq:nAnnulus}\\
\lambda_\mathrm{An} & = N\Bigl(\frac{4}{3}\pi (R_2^3-R_1^3)\Bigr)^{-1}.\nonumber
\end{align}We will refer to it as an annulus due to the shape of its cross section. They solved the steady state version of (\ref{eq:fEOM}) in all space by imposing continuity of velocity and shear stress at both the inner, $r=R_1$, and outer, $r=R_2$, surface. They found ($F_S = 6\,\pi\,\eta\,R_2\,U_0$, $\beta_i = R_i\sqrt{\gamma \lambda_\mathrm{An}/\eta},\,\, i=1,2$)
\begin{align}
F/F_S = &\frac{1}{3}\Bigl[(\cosh \beta_2 - \frac{\sinh \beta_2}{\beta_2})H_2(\beta_1,\beta_2)\nonumber\\
& - (\sinh \beta_2 - \frac{\cosh \beta_2}{\beta_2})H_1(\beta_1,\beta_2) \Bigr]\label{eq:ssAnnulusF},
\end{align}where the functions $H_1$ and $H_2$ can be found in \citet{BS93}. Moreover, the right-hand side of (\ref{eq:ssAnnulusF}) reduces to that of (\ref{eq:ssSphereF}) in the limit $\beta_1\rightarrow 0$ by identifying $\beta_2 = \beta$.

\subsection{Oscillating particle}

In this section, we study the hydrodynamic force ${\bf F}$ experienced by the infinitely thin shell and the uniform-density sphere oscillating in a fluid along a fixed axis. We briefly summarize recent work on the subject and present new time-dependent solutions to (\ref{eq:fEOM}).

\citet{S51} was the first to solve for the hydrodynamic force exerted on a solid sphere oscillating sinusoidally at an angular frequency $\omega$ in a quiescent incompressible fluid. His result and its generalizations are recapitulated by both \citet{L32} and \citet{LandauLifschitz87}. The velocity of the particle is assumed to be ${\bf v} = v \,\hat{{\bf e}}_z = v_0 e^{i\omega t}\,\hat{{\bf e}}_z$, which, due to symmetry considerations, allows one to describe the resulting fluid motion outside the particle as a doublet--stokeslet combination for which the stream function $\psi_O$ reads in spherical coordinates 
\begin{subequations}
\label{eq:stream}
\begin{align}
\psi_O & =  h_O(r)\,v\,\sin^2 \theta,\quad r \geq R;\label{eq:stream:a}\\
h_O(r) & =  A/r + (D/k)(1+1/(k\,r))e^{-k r},\label{eq:stream:b}
\end{align}
\end{subequations}where we have used the abbreviations $k=(1+ i)\alpha$ and $\alpha=\sqrt{\rho\omega / 2 \eta}$, and $h_O(r)$ contains only terms that vanish as $r \rightarrow \infty$. The polar angle $\theta$ is measured with respect to the $z$ axis, {\it i.e.} direction of particle or ambient flow velocity. Stokes solved the linearized equation (\ref{eq:fEOM}) by imposing a no-slip boundary condition on the surface of a solid sphere: ${\bf u}(\vert {\bf r} \vert = R,\theta) \equiv {\bf v}(\vert {\bf r} \vert = R,\theta)$, which corresponds to the limit $\gamma \rightarrow \infty$ in (\ref{eq:F}).
The imposition of impenetrability and the no-slip boundary condition on the (outer) surface of the sphere amounts to four equations for the complex coefficients $A$ and $D$ in (\ref{eq:stream}). For any of the particles of (\ref{eq:nSphere}), (\ref{eq:nShell}) or (\ref{eq:nAnnulus}), the coefficients $A$ and $D$ will in general be different due the node distribution inside $B(t)$, but when the coupling constant $\gamma$ in (\ref{eq:F}) goes to zero they should also go to zero for any value of $\alpha$. A suitable stream function for solving (\ref{eq:fEOM}) inside the sphere has the form~\citep{VS09}
\begin{subequations}
\label{eq:streaminside}
\begin{align}
\psi_I & =  h_I(r)\,v\,\sin^2 \theta,\quad r \leq R;\label{eq:streaminside:a}\\
h_I(r) & =  B r^2 + C \Bigl(\frac{\sinh(k_I r)}{k_I r} - \cosh (k_I r)\Bigr),\label{eq:streaminside:b}
\end{align}
\end{subequations}where $k_I R = \sqrt{\beta^2 + 2 i Y^2}$ and $Y=R\alpha$. If the Laplacian term of (\ref{eq:fEOM}) is neglected inside the sphere, we may set $C$ to zero in (\ref{eq:streaminside}).  We remark that the parameter $Y$ is related to the Womersley number~\citep{W55} by $W_o=\sqrt{2}Y$, which expresses the ratio of oscillatory fluid inertia to the shear force.

We may proceed to calculate the fluid velocity field components, $u_r$ and $u_\theta$, and components of the stress tensor, $\sigma_{r\theta}$ and $\sigma_{rr}$, in a given region of space via
\begin{align}
u_r & = (r^2 \sin \theta)^{-1} \partial_\theta \psi; \quad u_\theta = -(r\sin \theta)^{-1} \partial_r \psi;\label{eq:velocomponents}\\
\sigma_{r\theta} & = \eta\Bigl(\frac{1}{r}\frac{\partial u_r}{\partial \theta} + \frac{\partial u_\theta}{\partial r} - \frac{u_\theta}{r}\Bigr);\label{eq:stressrrcomp}\\
\sigma_{rr} & = -p + 2\eta\frac{\partial u_r}{\partial r}\label{eq:stressrtcomp}.
\end{align}The pressure, $p$, is associated with the irrotational part, {\it i.e.} $\psi_\mathrm{doublet} = (A/r) v \sin^2 \theta$ in (\ref{eq:stream:a}), of the flow and is solved from $\nabla p = -\rho \, \partial {\bf u}_\textrm{doublet} /\partial t$. We note that in determining the pressure distribution inside the sphere in the DBB model, the contribution due to the irrotational part of the force ${\bf f}$ must be included.

Once the stress tensor is known, (\ref{eq:hForce}) or (\ref{eq:F}) may be used to calculate the total hydrodynamic force on the sphere with arbitrary $A$ and $D$ as
\begin{equation}
{\bf F} = F\,\hat{{\bf e}}_z = - 2\,\pi\,\rho\,\omega\,v R^3 \Bigl[ \frac{i}{3}+\frac{3}{2}\frac{1+k\,R}{Y^2}\Bigr] \Omega\,\hat{{\bf e}}_z,\label{eq:Ffactoring}
\end{equation}where the dimensionless function $\Omega = \Omega_\textrm{Re} + i\Omega_\textrm{Im}$ reads
\begin{equation}
\Omega = R^{-3}\frac{4AiY^2 - 4 D R^2 e^{-kR}(1+(1+i)Y)}{9(1+(1+i)Y)+2iY^2}\label{eq:OmegaForm},
\end{equation}and $A$ and $D$ have units of length cubed and length, respectively. Equation~(\ref{eq:Ffactoring}) has the attractive feature that $\Omega = 1$ corresponds to the impermeable no-slip result for a sphere.  Setting $\Omega = 1$ and taking the real part reduces it to ($v = v_0 e^{i\omega t}$)
\begin{align}
\mathcal{R}\lbrace F \rbrace & = - 6\pi \eta R v_0 \cos (\omega t) (1+Y)\label{eq:dynFaStokes}\\
& - (4/3)\pi \rho R^3 (-\omega v_0) \sin (\omega t) \Bigl(\frac{1}{2}+\frac{9}{4 Y}\Bigr)\nonumber,
\end{align}which is the no-slip result valid for a solid sphere obtained by \citet{S51}, where $\mathcal{R}\lbrace F \rbrace$ denotes the real part of $F$.

For the more general case ($\Omega\neq 1$), $\mathcal{R}\lbrace F \rbrace$ can be written as
\begin{subequations}
\label{eq:dynF}
\begin{align}
\mathcal{R}\lbrace F \rbrace & = - 6\pi \eta R v_0 \cos (\omega t) C_\mathrm{s}\nonumber\\ 
&\phantom{=\,\,}-(4/3)\pi \rho R^3 (-\omega v_0) \sin(\omega t) C_\mathrm{Ad};\label{eq:dynFa}\\
C_\mathrm{s} & = \Bigl(1 + Y\Bigr)\Omega_\textrm{Re} - \Bigl(Y + \frac{2}{9}Y^2\Bigr)\Omega_\textrm{Im};\label{eq:dynFb}\\
C_\mathrm{Ad} & = \Bigl(\frac{1}{2}+\frac{9}{4 Y}\Bigr)\Omega_\textrm{Re}  + \Bigl(\frac{9}{4 Y} + \frac{9}{4 Y^2}\Bigr)\Omega_\textrm{Im}\label{eq:dynFc},
\end{align}
\end{subequations}where we have again used the fact that $\dot{v} = i\omega v$. The factors $C_s$ and $C_\textrm{Ad}$ depend both on the frequency and the boundary condition on the surface of the sphere (through $A$ and $D$). Having derived (\ref{eq:dynF}), Vainshtein and Shapiro attempted to generalize (\ref{eq:dynFaStokes}) to a uniform-density sphere (see (\ref{eq:nSphere})) by applying the boundary condition (\ref{eq:VSBC}) and using (\ref{eq:VSinside}) inside the sphere without the Laplacian term. They refer to this as the ``Darcy model'' to which their solution reads~\citep{VS09}
\begin{equation}
\label{eq:OOmegaVS}
\Omega_\textrm{VS} = \frac{2 \beta^2 + 4 i Y^2}{2\beta^2 + 3 + 3((1+i)Y + 2 i Y^2)},
\end{equation}where $\beta = R\sqrt{\gamma \lambda/\eta}$. Equation~(\ref{eq:OOmegaVS}) is, however, unsatisfactory since
$$\Omega_\textrm{VS} \rightarrow \frac{4 i Y^2}{3 + 3((1+i)Y + 2 i Y^2)}\,\,\textrm{when}\,\,\beta\rightarrow 0.$$
When $\beta\rightarrow 0$ the particle is completely permeable and the hydrodynamic force should vanish altogether (the fluid and particle phase become completely decoupled in the framework of the DBB model (\ref{eq:fEOM})). They also presented an approximate generalization for the Brinkman problem with the Laplacian term of (\ref{eq:fEOM}) inside $B(t)$ included, but it inherited the problem of (\ref{eq:OOmegaVS}). We may lift the discrepancy of (\ref{eq:OOmegaVS}) by considering the DBB model of (\ref{eq:fEOM}) without the Laplacian term inside $B(t)$. Thus, we set $C$=$0$ in (\ref{eq:streaminside:b}) and use (\ref{eq:velocomponents})-(\ref{eq:stressrtcomp}) to calculate the resulting velocity and pressure fields inside and outside $B(t)$ via (\ref{eq:streaminside}) and (\ref{eq:stream}), respectively. We require continuity of velocity and pressure on the surface ${\bf r}=R\hat{{\bf e}}_r$ as opposed to (\ref{eq:VSBC}). This formulation gives
\begin{equation}
\Omega_\textrm{D} = \frac{2 \beta^2}{2\beta^2 + 3 + 3((1+i)Y + 2 i Y^2)},\label{eq:omegaVScorr}
\end{equation}which features the limits $\Omega_D \rightarrow 0$ as $\beta \rightarrow 0$ and $\Omega_D \rightarrow 1$ as $\beta \rightarrow \infty$. The reason for the unphysical $\beta \rightarrow 0$ limit of (\ref{eq:OOmegaVS}) can thus be attributed to the boundary condition (\ref{eq:VSBC}), which does not correspond to continuity of fluid velocity at the boundary if ${\bf v}\neq 0$. The $Y\rightarrow 0$ limit of (\ref{eq:omegaVScorr}) agrees with (\ref{eq:ssSphereFapprox}). This now gives us an expression with physically reasonable limits, but only within the Darcy model approximation.

\citet{LC04} considered the hydrodynamic force experienced by a sinusoidally oscillating nearly impermeable sphere using a perturbative expansion in $1/\beta$. However, they assumed the normal component of the fluid velocity to be zero on the surface of the particle, {\it i.e.} ${\bf u} \cdot \hat{{\bf e}}_r = 0$ for ${\bf r} \in \partial B(t)$. They found the hydrodynamic force to be
\begin{equation}
\label{eq:fLC}
F/F_s = 1+k+k^2/9 - (1+k)^2/(\beta\xi),
\end{equation}where $\beta=\kappa R$ as defined in the present work and $\xi$ is a slip coefficient related to the tangential fluid slip length $\Xi$ by $\Xi=1/(\kappa\xi)=R/(\beta \xi)$~\citep{LC04}. We may compare their model to ours by using (\ref{eq:dynF}) to extract the coefficients $C_s$ and $C_\textrm{Ad}$, which become
\begin{subequations}
\label{eq:LCF}
\begin{align}
C_s & = 1 + Y - (1+2Y)\frac{\Xi}{R};\label{eq:LCF:a}\\
C_\mathrm{Ad} & = \frac{1}{2} + \frac{9}{4 Y} - \frac{9}{2}\Bigl(1+\frac{1}{Y}\Bigr)\frac{\Xi}{R}\label{eq:LCF:b}.
\end{align}
\end{subequations}We note that within the DBB model, we do not concern ourselves with a decomposition of the hydrodynamic coupling, {\it i.e.} $\gamma$, to a normal and a tangential component. However, in order to compare Looker and Carnie's result to ours, we need to determine a value for $\xi$. In keeping with their theory, we have assumed $\xi$ to be a fixed number, which we have determined to be $\xi=0.9$ by least squares fitting the $Y\rightarrow 0$ limit of (\ref{eq:LCF:a}) to (\ref{eq:ssSphereF}) in the range $\beta \in [5,100]$. We will use this value of $\xi$ throughout the paper as a similar fit to (\ref{eq:ssSphereFapprox}) gives poorer agreement. One immediately sees that the factors in (\ref{eq:LCF}) reduce to those in (\ref{eq:dynFaStokes}) as the slip on the surface vanishes $\Xi \rightarrow 0 \Leftrightarrow \beta \rightarrow \infty$, but due to the nature of the expansion, the weak-coupling limit is not correctly reproduced. We comment more on (\ref{eq:LCF}) in the Results.

\subsection{Exact solution for the oscillating shell}\label{sec:exactshell}
Finding full non-perturbative solutions to the time-dependent equations (\ref{eq:fEOM}) with general force densities like those in (\ref{eq:forcedensity}), not just using the Darcy approximation, has not been done.  In particular, there does not exist an analytical solution to (\ref{eq:fEOM}) for the shell of (\ref{eq:nShell}) oscillating in a viscous fluid. We establish such a solution here by requiring a matching condition between the force on the shell determined both by the coupling (\ref{eq:forcedensity}) in the DBB model and the fluid stress (\ref{eq:hForce}):
\begin{align}
\int_{B(t)} -\gamma n_\textrm{Sh}({\bf r}) ({\bf v} - {\bf u})\,\mathrm{d}^3 x & = \int_{\partial B(t)} \hat{{\bf e}}_r \cdot \boldsymbol{\sigma}\,\mathrm{d}S,\label{eq:Fbalance}
\end{align}where the integrand on the left-hand side is a force per volume. Equation~(\ref{eq:Fbalance}) has sound physical limits as it guarantees there will be no force acting on the particle if $\gamma = 0$ since in that case the only solution is to have $A=D=0$. Integrating over $r$ changes the left-hand side into a surface integral and we may equate the integrands, {\it i.e.}
\begin{align}
-\gamma \lambda_\mathrm{Sh}({\bf v}-{\bf u}) = \hat{{\bf e}}_r \cdot \boldsymbol{\sigma}\label{eq:localFbalance}\\
\Leftrightarrow \left \{
\begin{array}{ll}
 -\gamma\lambda_\textrm{Sh}(v_r - u_r) & = \sigma_{rr};\\
 -\gamma\lambda_\textrm{Sh}(v_\theta - u_\theta) & = \sigma_{r\theta}
 \end{array}\right.\nonumber
\end{align}is required to hold for $0 \leq \theta \leq \pi$ at $r=R$ at any given time; ${\bf u}$ and $\boldsymbol{\sigma}$ are based on the stream function of (\ref{eq:stream:b}). By solving (\ref{eq:localFbalance}), we obtain closed-form expressions for the unknowns $A=A_\textrm{Re}+i A_\textrm{Im}$ and $D=D_\textrm{Re} + i D_\textrm{Im}$ in terms of $\beta$ and $Y$. We note that the algebra involved is simplified by the substitution $D=\tilde{D}\exp{k R}$ and by the use of a symbolic computation package like Mathematica. Plugging the resulting expressions into (\ref{eq:OmegaForm}) gives the real and imaginary parts of (\ref{eq:OmegaForm}) for the shell as 
 $\Omega_\textrm{Re} = \Omega_1/\Omega_3$ and $\Omega_\textrm{Im} = \Omega_2/\Omega_3$, where
\begin{subequations}
\label{eq:ourOmega}
\begin{align}
\Omega_1 & = \beta^8 \Bigl( 81 + 162 Y + 162 Y^2 + 36 Y^3 + 4 Y^4 \Bigr)\nonumber\\ 
& + (9/2) \beta^6 \Bigl(729 + 1539 Y + 1620 Y^2 + 486 Y^3 + 80 Y^4 + 12 Y^5 \Bigr)\nonumber\\
& + 18 \beta^4 \Bigl(2187 + 5103 Y + 5832 Y^2 + 2430 Y^3 + 504 Y^4 + 144 Y^5 + 54 Y^6 + 
 4 Y^7 \Bigr)\nonumber\\
& + 1458 \beta^2 (1 + Y) \Bigl(81 + 162 Y + 162 Y^2 + 36 Y^3 + 9 Y^4 + 6 Y^5 + 2 Y^6 \Bigr);\\
\Omega_2 & = -Y \Biggl((3/2) \beta^6 (9 + 2 Y) \Bigl(27 + 54 Y + 54 Y^2 + 12 Y^3 + 4 Y^4 \Bigr)\nonumber\\
& + 18 \beta^4 \Bigl(729 + 2 Y (810 + 891 Y + 324 Y^2 + 81 Y^3 + 21 Y^4 + 4 Y^5)\Bigr)\nonumber\\
& + 162 \beta^2 (9 + 2 Y) \Bigl(81 + 162 Y + 162 Y^2 + 36 Y^3 + 9 Y^4 + 6 Y^5 + 2 Y^6 \Bigr)\Biggr);\\
\Omega_3  & = \Bigl(81 + 162 Y + 162 Y^2 + 36 Y^3 + 4 Y^4\Bigr)\Bigl(\beta^8 + 9 \beta^6 (5 + Y) \nonumber\\
& + (9/4) \beta^4 (297 + 162 Y + 18 Y^2 + 4 Y^3 + 4 Y^4)\nonumber\\
& + 27 \beta^2 (135 + 162 Y + 54 Y^2 + 12 Y^3 + 6 Y^4 + 2 Y^5) \nonumber\\
& + 81 (81 + 162 Y + 162 Y^2 + 36 Y^3 + 9 Y^4 + 6 Y^5 + 2 Y^6) \Bigr),
\end{align}
\end{subequations}which is now an exact result for the shell.

The relevant limits of (\ref{eq:ourOmega}) merit comment. The zero-frequency limit ($Y\rightarrow 0$) for our model gives 
$$F/F_S \rightarrow 2\beta^2/(2\beta^2 + 9)$$ and 
$$C_\mathrm{Ad} \rightarrow 0.$$ 
This is consistent with performing the force matching using the zero-frequency stream functions that yields the steady-state shell results of (\ref{eq:ssShell}). The infinite-frequency limit gives 
$$\lim_{\omega \rightarrow \infty}  F/F_S = \lim_{\omega \rightarrow \infty} C_s = 2 \beta^2/9$$ and 
$$\lim_{\omega \rightarrow \infty} \omega C_\mathrm{Ad} = 0.$$ 
If we allow $\beta$ to tend to infinity, these four limits agree with those for the no-slip boundary condition.
For the infinite-coupling limit ($\beta \rightarrow \infty$, $Y$ given), we find $$C_s \rightarrow 1 + Y$$ and $$C_\mathrm{Ad} \rightarrow 1/2 + 9/(4Y),$$ which agree with Stokes's no-slip result of (\ref{eq:dynFaStokes}).  Thus, (\ref{eq:ourOmega}) matches up with all the known relevant limits.

\subsection{Exact solution to the oscillating uniform-density sphere}\label{sec:exactsphere}
Another case for which there exists no analytical solution to the time-dependent equation (\ref{eq:fEOM}) is that of the uniform-density sphere of (\ref{eq:nSphere}) oscillating in the fluid with velocity ${\bf v} = v_0 e^{i \omega t} \hat{{\bf e}}_z$. We have solved this problem by formulating the system of equations (the fields based on (\ref{eq:stream})/(\ref{eq:streaminside}) are denoted with/without the tilde),
\begin{equation}
\label{eq:exactSphere}
\tilde{u}_r = u_r, \quad \tilde{u}_\theta = u_\theta; \quad \tilde{\sigma}_{rr} = \sigma_{rr}, \quad \tilde{\sigma}_{r\theta} = \sigma_{r\theta}, \forall\,\,{\bf r} \in \partial B(t)
\end{equation}which are eight equations for the real and imaginary components of the four complex numbers $A$, $B$, $C$ and $D$ of (\ref{eq:stream:b}) and (\ref{eq:streaminside:b}). 
The use of a symbolic computation package vastly simplifies the task of deriving these results.  The resulting real and imaginary parts $\Omega_\textrm{Re} = \Omega_1/\Omega_3$ and $\Omega_\textrm{Im} = \Omega_2/\Omega_3$ are
\begin{subequations}
\label{eq:ourOmegaForSphere}
\begin{align}
\Omega_1 & = 2\beta^2 \Bigl( P_1 \cos (2 Y^2/Z) - 2 Z Y P_2 \sin (2 Y^2/Z) \nonumber\\
& + P_3 \cosh (2 Z) + 2 Z P_4 \sinh(2 Z) \Bigr);\\
\Omega_2 & = -6\beta^2 \Bigl( Y P_5 \cos (2 Y^2/Z) - 2 Z P_6 \sin (2 Y^2/Z) \nonumber\\
& + Y P_7 \cosh (2 Z) + 2 Z Y P_8 \sinh (2 Z) \Bigr);\\
\Omega_3 & = \Bigl(81 + 162 Y + 162 Y^2 + 36 Y^3 + 4 Y^4\Bigr)\nonumber\\
& \Bigl((Z^2-Y^2)\bigl( P_9 \cos(2 Y^2/Z) - 2 Z Y P_{10} \sin(2 Y^2/Z)\bigr) \nonumber\\
& +(Z^2 + Y^2) \bigl(P_{11} \cosh(2 Z)+ 2 Z P_{12} \sinh (2 Z)\bigr)\Bigr),
\end{align}
\end{subequations}where $\sqrt{2}Z=\sqrt{\beta^2+\sqrt{\beta^4+4Y^4}}$ and the $P_j=P_j(\beta,Y)$ are polynomials that are provided in Appendix \ref{appendixb}.

The result(\ref{eq:ourOmegaForSphere}) has the low frequency, $Y\rightarrow 0$, limit 
$$ F/F_S \rightarrow  \frac{2\beta^2(1 - \beta^{-1}\tanh \beta)}{2\beta^2 + 3 (1 - \beta^{-1}\tanh \beta)}$$
and 
$$ C_\textrm{Ad} \rightarrow 0, $$ 
which agrees with the steady-state result of Debye and Bueche (\ref{eq:ssSphereF}).
The $\beta \rightarrow \infty$ limits (impenetrable, no-slip limit) are $C_s \rightarrow 1+Y$ and $C_\textrm{Ad} \rightarrow 1/2 + 9/(4 Y)$, in agreement with Stokes's no-slip result of (\ref{eq:dynFaStokes}).

\section{Results: Comparison to simulations}

In this section, we compare theoretical predictions to Lattice-Boltzmann simulations for various node densities, for both steady state and as the particle oscillates sinusoidally.

\subsection{Steady-state}

\begin{figure}
\centerline{\includegraphics[width=0.55\textwidth]{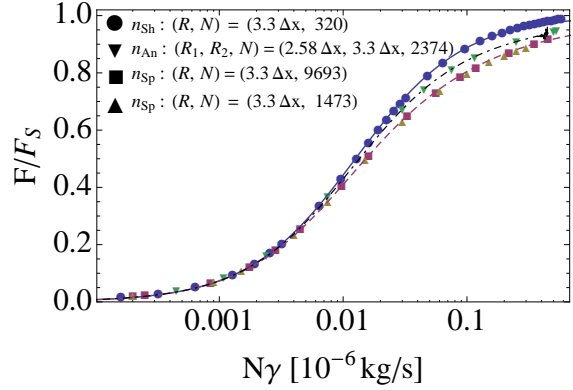}}
\caption{Comparison of normalized steady-state drag force $F/F_S$ versus $N\gamma$ for particles with different node density distributions $n({\bf r})$. The lines are the theoretical predictions for the shell ((\ref{eq:ssShellF}), solid line), uniform-density sphere ((\ref{eq:ssSphereF}), dashed line) and the annulus ((\ref{eq:ssAnnulusF}), dot-dashed line). Taking the thin-annulus limit of (\ref{eq:ssAnnulusF}) leads to numerical cancellation errors for large $N\gamma$, but the agreement with simulations is still good.}
\label{fig:figure1}
\end{figure}

\begin{figure}
\centerline{\includegraphics[width=0.55\textwidth]{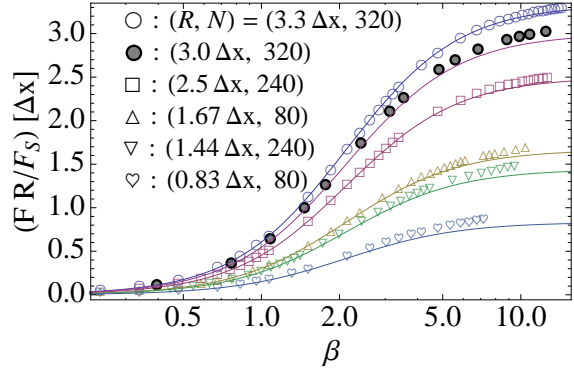}}
\caption{Normalized steady-state drag force for a shell multiplied by the radius $R$ at which the nodes sit. As the radius is increased to $R=3.3\Delta x$, the simulation results differ from the theory by no more than $0.6\%$ at $\beta = 14.4$.}
\label{fig:figure2}
\end{figure}

In figure~\ref{fig:figure1}, we compare theoretical predictions to LB simulations for the steady-state drag force on the shell, (\ref{eq:ssShellF}), uniform-density sphere, (\ref{eq:ssSphereF}), and on the annulus, (\ref{eq:ssAnnulusF}). The results are presented as a function of the product $N \gamma$ of the node number and coupling parameter as one cannot characterize the annulus in terms of a single $\beta$ that is proportional to the radius. The measurement is performed in a cubic simulation box with periodic boundary conditions in $x$ and $z$ directions and parallel no-slip walls at $y=0$ and $y=30\,\Delta x$ moving at velocity ${\bf u}_\infty = U_0\,\hat{{\bf e}}_z = (0.0001\Delta x/\Delta t)\,\hat{{\bf e}}_z$. The particle is kept immobile at the centre of the box. We choose this arrangement as it reduces finite-size effects~\citep{LM76} in the velocity field at least to $1/L^2$ from $1/L$.

We find our simulation results to be in excellent agreement with the theoretical predictions without any fitting parameters. The simulation method is therefore able to differentiate between a shell, an annulus and a uniform-density sphere without ambiguity. It is noteworthy that even a sub-grid thick annulus ($R_2-R_1 = 0.72\Delta x$) is so closely matched with the theoretical prediction.  This is most likely due to the fact that the node has a compact support on the fluid mesh which in this case couples a node only to the unit cell in which it resides~\citep{AD98,SD07,OSD11}.

Figure~\ref{fig:figure1} is, however, a more ideal case for the simulations where $R/\Delta x = 3.3$ was not commensurate with the underlying fluid mesh.  In figure~\ref{fig:figure2}, we compare shells of different radius -- node count combinations to the prediction of (\ref{eq:ssShellF}). We observe the simulation results to agree with the prediction for all $\beta = R\sqrt{\gamma \lambda/\eta}$ in the case of the largest $R=3.3\Delta x$. The only caveat is the apparent mismatch between theory and simulation for the smaller shells at large $\beta$. A change in the surface area per node, $4\pi R^2/N$, did not reduce the mismatch, for which reason we ran the simulation for different  radius -- node count combinations. This suggests that the mismatch is purely a discretization effect of immersing an off-lattice spherical shell into a cubic lattice. Figure~\ref{fig:figure2} does not just highlight the effect of increasing particle size, but it also indicates the importance of incommensurability with the lattice. The two largest particles are nearly of the same size, but for the commensurate particle $R/\Delta x = 3$, the lattice effects are emphasized as $\beta$ increases compared to the incommensurate case $R/\Delta x = 3.3$ and the lattice effects again decrease for $R/\Delta x = 2.5$.  This commensurability effect makes it difficult to calculate analytically the discretization errors from this sort of immersed boundary simulation.  However, if we fit the analytical curves to the simulation data by allowing the radius $R$ to be a fitting parameter, then we find that the fitted radius differs from the radius at which the discrete nodes sit by no more than $0.1\,\Delta x$ for the cases plotted in figure~\ref{fig:figure2}, thus putting a bound on the discretization errors.

\begin{figure}
\begin{center}$
\begin{array}{cc}
\includegraphics[width=0.48 \textwidth]{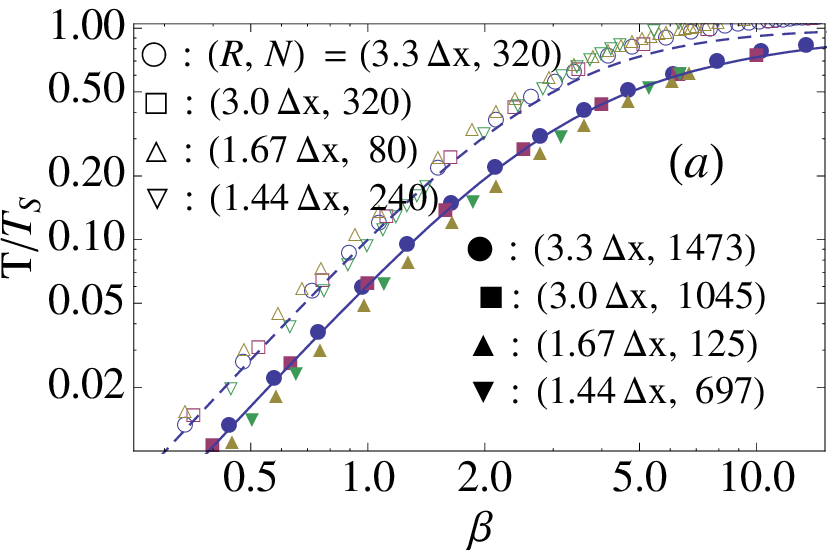}&
\includegraphics[width=0.48 \textwidth]{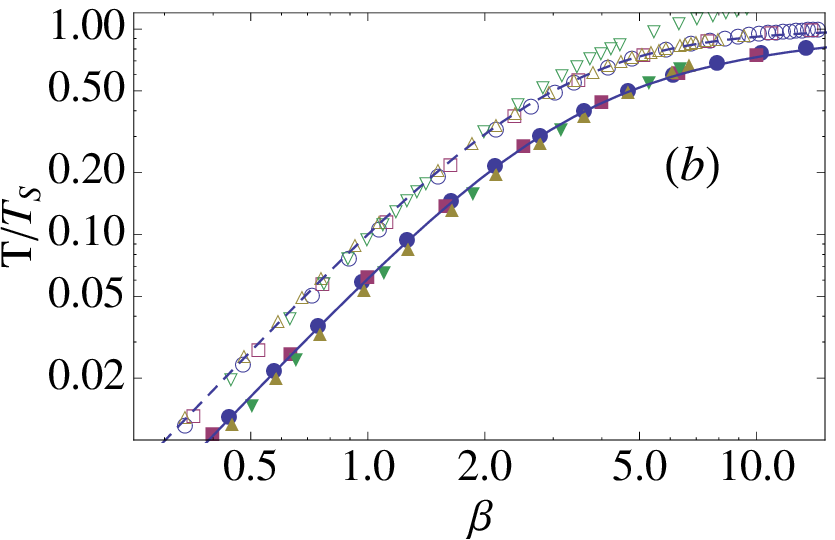}
\end{array}$
\end{center}
\caption{Normalized steady-state drag torque on uniform-density spheres (solid symbols) and infinitely thin shells (hollow symbols) of different radii and node counts. (a) Measurement data are shown without any fitting. (b) Normalized measurements are shown with $\Delta R$ in $T_S=4\,\pi \eta (R+\Delta R)^3 Q$ used as the fitting parameter, which falls between $\Delta R = 0.10\,\Delta x$ and $0.15\,\Delta x$.}
\label{fig:figure3}
\end{figure}

The drag torque experienced by the uniform-density sphere, (\ref{eq:ssSphereT}), and by the shell, (\ref{eq:ssShellT}), is more sensitive to changes in the radius $R$ than Stokes drag as the dependence is cubic. We have measured the drag torque on spheres and shells of radii $3.3\,\Delta x$, $3.0\,\Delta x$, $1.67\,\Delta x$ and $1.44\,\Delta x$. We present the measurements together with theoretical predictions in figures~\ref{fig:figure3}(a) and (b). Figure~\ref{fig:figure3}(a) shows the theory without any fitting parameters whereas in figure~\ref{fig:figure3}(b), the radius has been used as the fitting parameter. Even though the fitting parameter turns out to be a very small effect, giving a fitted radius $0.1\,\Delta x$ larger than the radius at which the nodes sit, it has a noticeable impact as the torque is proportional to $R^3$. Moreover, the effect of the compact support of the nodes on the fluid mesh has a big effect on small ($R\approx \Delta x$) spheres since its support spreads the nodes out to radii between $\Delta x$ and $2\Delta x$.  As with the drag force, the discretization errors for the torque can be minimized by using an irrational ratio $R/\Delta x$, for which the particle is incommensurate with the lattice~\citep{OSD11}.

\subsubsection{The limit of impermeability}
We conclude the discussion of the steady-state results by referring to the limit of an impermeable particle, {\it i.e.} when $F/F_S = T/T_S = 1$. In the context of simulations, we refer to this limit as saturating $\beta$ since it requires $\gamma$ to be increased until impermeability is reached numerically. One might think that one could simulate a porous particle and consider it to be equivalent to an impermeable particle with a smaller effective ``hydrodynamic'' radius.  However, based on (\ref{eq:ssShellF}) and (\ref{eq:ssShellT}), one cannot simulate nodes placed at radius $R$ for a sub-saturation $\beta$ and claim to be modelling a particle with an effective radius $\tilde{R}$ based on, say, Stokes drag ${\bf F} = 6\pi\eta \tilde{R} {\bf v}$ as a measurement of the drag torque will give a very different value for the effective radius, which can be seen, for instance, by writing (\ref{eq:ssShellT}) as
\begin{displaymath}
T = 4 \pi \eta Q \bar{R}^3 \equiv 4 \pi \eta Q \Bigl(R \beta^{2/3}/(\beta^2+9)^{1/3} \Bigr)^3.
\end{displaymath}Clearly the ``effective'' hydrodynamic radius from the torque $\bar{R}$ and from Stokes drag, $\tilde{R} = 2\beta^2 R/(2\beta^2 + 9)$, are very different numbers unless $\beta$ is large. Therefore, in simulating an impenetrable sphere, one must increase $\beta$ until the hydrodynamic radii based on different measures all agree to within an acceptable level of tolerance. If one chooses a value for $\beta$ below saturation, one may only claim to be simulating a porous particle of the chosen node density $n({\bf r})$.  That is, one should be wary of models using an effective hydrodynamic radius as there is more than one way to define such a radius and different measures will not generally give the same measure. The effective-radius concept has also been criticized by~\citet{ACE-JNW10b}.

In theory, the limit of impermeability requires $\beta\rightarrow\infty$.  However, it is clear from figures~\ref{fig:figure2} and \ref{fig:figure3} that the drag force and torque are typically slightly larger in simulation at large $\beta$ than theory suggests due to discreteness of simulations.  Thus, the limit of impermeability can be reached in practice in actual simulations for a finite, but still quite large, value of $\beta$.  Based on figures~\ref{fig:figure2} and \ref{fig:figure3}, particles of different radii and node distributions reach the limit at different values of $\beta$. The approach to the limit is dictated by the node density and the ratio $R/\Delta x$.  A slightly modified simulation algorithm may also be needed for stability at large values of $\beta$.~\citep{OSD11}

It is also clear from figure~\ref{fig:figure2} that for a given number of nodes $N$, a shell will approach the impermeable limit at a lower value of $\gamma$ than a uniform density sphere, which might make it preferable in some simulations.

\subsection{Oscillating particle}

\begin{figure}
\centerline{\includegraphics[width=0.45\textwidth]{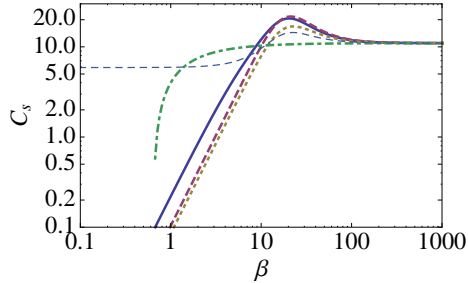}}
\caption{The correction factor $C_s$ plotted at $Y=\alpha R=10$ for our new shell theory ((\ref{eq:ourOmega}), solid line), the corrected Darcy model ((\ref{eq:omegaVScorr}), heavy dashed line), our new sphere theory ((\ref{eq:ourOmegaForSphere}), dotted line) and Looker and Carnie's perturbative expansion ((\ref{eq:LCF}), dot-dashed line), and Vainshtein and Shapiro's expression ((\ref{eq:OOmegaVS}), thin dashed line). Our two models new to this work and the corrected Darcy result have the correct asymptotic zero and infinite $\beta$ limits.}
\label{fig:figure4}
\end{figure}

We first compare the different theoretical predictions for the oscillating sphere case.  The steady-state uniform-density sphere result for the force, (\ref{eq:ssSphereF}) and (\ref{eq:ssSphereFapprox}), and the shell result, (\ref{eq:ssShellF}), are qualitatively very similar. In particular, both the steady-state drag force and torque go to zero as $\beta$ goes to zero. One would therefore expect similar behaviour from the oscillatory force on both the sphere and the shell. Figure~\ref{fig:figure4} shows the correction factor $C_s$ to the velocity-dependent part of the hydrodynamic drag force (\ref{eq:dynFa}) at $Y$=$\alpha R$=$R\sqrt{\rho \omega/(2\eta)}$=$10$ as a function of the dimensionless coupling parameter $\beta$ for Vainshtein and Shapiro's Darcy model (\ref{eq:OOmegaVS}), the corrected Darcy model (\ref{eq:omegaVScorr}), Looker and Carnie's nearly impermeable sphere (\ref{eq:LCF}) and our shell (\ref{eq:ourOmega}) and sphere model (\ref{eq:ourOmegaForSphere}). Our theories (solid and dotted lines) and the corrected Darcy model capture the correct approach to the limit $\beta \rightarrow 0$ in which the hydrodynamic force must vanish. We do emphasize that the existing theories are targeted to model a uniform-density sphere, but both the sphere and shell should have similar qualitative behaviour. It is interesting that Looker and Carnie's model based on homogenization theory does not lead to the peak in $C_s$ between $\beta=10$ and $100$ present in other theories.

\begin{figure}
\centerline{\includegraphics[width=\textwidth]{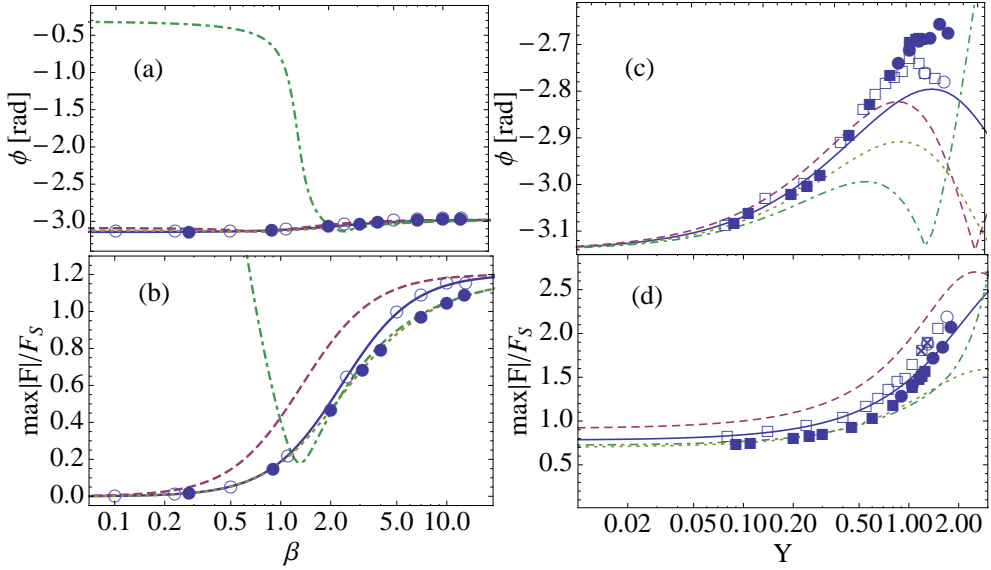}}
\caption{(a) Phase shift and (b) normalized amplitude of the oscillating hydrodynamic force as a function of $\beta$ at $Y=0.187$ (or, an angular frequency of $\omega$ = $2\pi/(5000\Delta t)=1.25\,\mathrm{MHz}$). The hollow circles correspond to simulations of a shell $(R,N)$=$(3.3\Delta x$=$330\,{\rm nm}, 540)$ with $\Delta x=100$ nm.  The data for a uniform-density sphere $(R,N)$=$(3.3\Delta x,2247)$ are plotted as solid circles. (c) Phase shift and (d) normalized amplitude of the oscillating hydrodynamic force as a function of $Y$ ($\omega$ was varied to change $Y$) at fixed $\beta=4.0$. Simulations in (c) and (d) were performed with a finer mesh resolution, $\Delta x=77\,\mathrm{nm}$ and time step $\Delta t=0.59\,{\rm ns}$. The hollow squares and circles correspond to simulations of shells $(R,N)$=$(4.3\Delta x, 540)$ and $(9.1\Delta x,2252)$. The data for uniform-density spheres $(R,N)$=$(4.3\Delta x,5665)$ and $(9.1\Delta x,61805)$ are plotted as solid squares and circles, respectively.
In all plots, lines correspond to our shell theory (solid line), corrected Darcy theory (dashed), our theory for the uniform-density sphere (dotted) and Looker and Carnie's perturbation theory (dot-dashed) with $\xi=0.9$.}
\label{fig:figure5}
\end{figure}

We now compare directly the predictions of the different time-dependent models, after substitution into (\ref{eq:dynF}), to simulation.  In the simulation we set the velocity of the rigid test particle to be equal to ${\bf v}(t) = v_0 \cos(\omega t)\hat{{\bf e}}_z$ and calculated the force exerted on it.  We placed the test particle at the centre ${\bf X}_0$ of the simulation box of size $10R\times 10R\times 20R$ and imposed periodic boundary condition in the $x$ and $z$ directions and had parallel no-slip walls in the $y$ direction to reduce finite-size effects~\citep{LM76}. The particle's velocity can be integrated analytically so we set its position to ${\bf X}(t) = {\bf X}_0 + (v_0/\omega) \sin(\omega t))\hat{{\bf e}}_z$.  The length scale $1/\alpha$ is a boundary layer thickness associated with the acceleration of the fluid next to the particle. As $Y\rightarrow (R/\Delta x)$, $1/\alpha$ approaches $\Delta x$ and viscous effects are confined to a narrow boundary layer on the particle surface. Moreover, $2\Delta x$ is the smallest length scale in the computational method for which central-difference velocity gradients can be computed and thus, we expect deviations between theory and simulation due to discreteness of the mesh for $Y\geq (R/2\Delta x)$.  So, to explore large $Y$, we need finer resolution.

The hydrodynamic force on the particle is characterized by its amplitude $\max \vert F \vert$ and phase shift $\phi \equiv \lim_{t \rightarrow \infty} \arg(F(t)) - \arg(v(t))$. These two quantities are easy to evaluate since all the linear theories for the force we compare here can be written as
\begin{eqnarray}
F(t) & = &\sqrt{a^2+b^2}\cos(\omega t + \phi),\\
a & =& - (6\pi\eta R v_0) C_s; \qquad
b  = (6\pi\eta R v_0) (4/9)Y^2 C_\mathrm{Ad};\nonumber\\
\phi & = &-\arccos \Bigl(a/\sqrt{a^2+b^2}\Bigr),\nonumber
\end{eqnarray}
for $v(t) = v_0 \cos(\omega t)$.  Since we have available theories for a uniform-density sphere and the shell, we have simulated both types of particles. The results for the simulations and different theoretical models are shown in figure~\ref{fig:figure5}.  As with the steady state results, we see that the simulation results for the uniform-density sphere and the shell differ only at moderate to high values of $\beta$ and the amplitude of the force vanishes in both cases as $\beta\rightarrow 0$.

We look first at the phase difference between the set velocity ${\bf v}$ and the measured hydrodynamic force ${\bf F}$ as a function of $\beta=R \kappa$ for the angular frequency $\omega = 2\,\pi/(5000\,\Delta t)$. This translates to $Y=0.187$ with $R=3.33\Delta x$ and $1/\alpha = 17.8\Delta x$ for which inertial effects in the fluid and compressibility effects at the scale of the particle should be negligible, but oscillatory effects should still be clearly visible for large $\beta$. We have measured the phase shift from simulations and plot the results together with theoretical predictions in figure~\ref{fig:figure5}(a). The velocity amplitude was set to $v_0 = 10^{-5} v_c$.
In steady state, the drag force is opposite to the particle velocity, which means there is a phase shift of $-\pi$ radians between the two. This is found to carry over to the oscillatory case for small $\beta$. However, as $\beta$ increases, more of the fluid around the particle is dragged along with it ({\it i.e.} its virtual mass increases) and the phase shift drops somewhat. We find all theoretical models to give similar results for the phase shift at large $\beta$ corresponding to a nearly impermeable particle. However, for a highly porous particle (small $\beta$ corresponding to $\kappa^{-1}>R$), Looker and Carnie's theory breaks down as expected. Our new model based on force matching on the surface agrees with simulation results at all values of $\beta$ for both the shell (hollow circles, $(R,N)$=$(3.3\Delta x,540)$) and a uniform-density sphere (solid circles, $(R,N)$=$(3.3\Delta x,2247)$). The nodes inside the uniform-density sphere were placed at intervals of $0.41\,\Delta x$. We may therefore conclude that the phase shift is determined largely by the drag exerted on the surface for the different types of node distributions. 

Figure~\ref{fig:figure5}(b) shows the corresponding data for the normalized amplitude of the oscillatory force. For $\beta > 5$, our exact solution to the uniform-density sphere agrees with Looker and Carnie's model (\ref{eq:LCF}) with $\xi=0.9$. Our method of fitting their slip coefficient $\xi$ thus works well for large $\beta$ providing a mapping between the DBB model and the solution obtained via the homogenization procedure. Moreover, the Darcy-like model predicts a larger force amplitude than the Brinkman-like models, which is consistent with the full Brinkman solution having the reduction factor (\ref{eq:reductionfactor}) whose value is between zero and unity. The difference between the uniform-density sphere and shell for the simulations is also consistent with the results for the steady-state drag force which showed a similar difference in figure~\ref{fig:figure1}. As $\beta$ approaches zero, the Darcy model of (\ref{eq:omegaVScorr}) corrects the discrepancy of (\ref{eq:OOmegaVS}) that predicts a finite hydrodynamic force on the particle, which is unphysical. Our new models of sections \ref{sec:exactshell} and \ref{sec:exactsphere} based on local force matching agree well with the simulation data for all $\beta$. As in the steady-state case, the sphere and shell results coincide as $\beta\rightarrow 0$.

Last, we scan over the other parameter, $Y=R\alpha$, in the model for a fixed value of $\beta=R\sqrt{\gamma n/\eta}$ to see how the phase shift and normalized force amplitude are affected.  We could perform simulations for different $R$ such that fixed value $Y$ would be equivalent to different values of $\omega$. However, this should not make a difference to the force amplitude normalized by $F_S$. Alternatively, if we change $Y$ by adjusting the shear viscosity $\eta$, we must also change either $\gamma$ or $\lambda$ (or both) to keep $\beta$ unchanged.  In addition, a large increase in the shear viscosity can make a filled shell effectively a uniform-density sphere with a coating.  In the end, we fix a few values of $R$ and vary $\omega$ to change $Y$.   We pick an intermediate value of $\beta = 4.0$ for which different systems (shell versus spheres) and theories are clearly distinguished in figure~\ref{fig:figure5}(b). A finer resolution, $\Delta x=77\,{\rm nm}$ and time step $\Delta t=0.59\,{\rm ns}$ was used for these simulations. This was necessary as the length scale $1/\alpha$ that appears in (\ref{eq:stream}) becomes small in lattice units at higher frequencies for the original parameters. The length scale $1/\alpha$ sets the wavelength and decay length of the disturbance caused by the moving particle and was too short to be accurately resolved on the original mesh. The measurement of the phase turned out to be very sensitive to the box size for which reason we used a mesh as large as $12R\times 16R \times 28R$. In figure~\ref{fig:figure5}(c) we expectedly find all models to give reasonable phase shifts close to $\phi = -\pi$ at low frequencies ($Y<0.1$). As $Y$ is increased, the phase shifts determined from simulations both for the shell (hollow symbols) and the sphere (solid symbols) follow the model results well for $Y<1$.  Again, the phase shift for the shell and sphere are very similar and thus appears to be determined on the surface of the particle.

The normalized force amplitude in figure~\ref{fig:figure5}(d) confirms the findings of figure~\ref{fig:figure5}(b) both by theory and simulation: the uniform-density sphere (filled symbols) experiences a smaller drag force than the shell (hollow symbols). The simulation data for the uniform-density sphere agree well for $Y<0.8$ with our theory for the uniform-density sphere (dotted line) and with Looker and Carnie's theory with $\xi = 0.9$ (dot-dashed line) (as seen for $Y=0.187$ and $\beta \geq 3$ in \ref{fig:figure5}(b)). The shape of $F/F_S$ as a function of $Y$ for the shell, however, is captured quantitatively only by our new shell model for $Y<1.1$. The simulation data indicate a larger force on the particle than the theory roughly when $Y \geq 1$.  Doubling any of the dimensions of the already large simulation volume did not change the results either for which reason finite-size effects are not responsible for the difference either. We can therefore tentatively suggest that assumptions of the near field in linear Stokes theory break down when $1/\alpha \leq R$. We therefore restrict the regime of validity of the present theories to $Y<1.0$. It appears that Looker and Carnie's perturbative treatment maps closely to our uniform-density sphere model after fitting of $\xi$ for $Y<1.0$ even for $\beta=4$. The effect of fitting becomes less important as $\beta$ is increased, a result that is similar to what was stated in \citet{VS09}. A potentially important practical finding is that for $\beta=4$, force matching on the surface reasonably captures both the sphere and the shell measurements. The corrected Darcy model (dashed line) does have the correct qualitative shape as a function of $Y$ compared to the other models and our simulations, but it predicts too large a hydrodynamic force.

\section{Conclusions}
We have validated theoretical predictions of the steady-state drag force and drag torque on porous shells of different thickness and uniform-density spheres using Lattice-Boltzmann simulations of the full Navier-Stokes equations. We have found the drag force to be in quantitative agreement with the theory without any adjustable fitting parameters. The torque measurement proved to be more sensitive to discretization effects and commensurability between the particle and the underlying fluid mesh.

We derived new closed-form expressions for the hydrodynamic force on an oscillating shell and on a uniform-density sphere as a function of the coupling to the fluid. Our approach is in good quantitative agreement with simulations and it is consistent for all degrees of particle porosity, which is an improvement over existing analytical models. We have demonstrated that our models are able to predict the hydrodynamic force correctly when the porosity or the frequency of oscillation are changed independently. We have also pointed out the regime of validity of these theories that base themselves on the linear Stokes equation.

\begin{acknowledgments}
This work was supported  in part by the Academy of Finland through its COMP CoE grant, by the Natural Science and Engineering Council of Canada and by SharcNet. We also wish to thank CSC, the Finnish IT centre for science, for allocation of computer resources. S.T.T.O. would like to thank the National Graduate School in Nanoscience, the Finnish Cultural Foundation and the Alfred Kordelin Foundation for funding.
\end{acknowledgments}

\appendix
\section{LB method}
\label{appendixLB}
Lattice Boltzmann (LB) is an increasingly common scheme to solve the Navier-Stokes equations~\citep{S01,ST05} which we summarize here. The method is based on solving an approximation of the Boltzmann transport equation (BE) on a cubic mesh (or grid) with sites $\mathbf{x} = (i,j,k)\Delta x$ connected to their neighboring sites by a set of $n$ vectors $\{\mathbf{e}_i\}_{i=0}^{n-1}$ along which material is transported according to a discretized version of the BE. We define a distribution function $g_i({\bf x},t)$ where $i$ labels the lattice directions from site ${\bf x}$. For three-dimensional systems, we use a 15-velocity model~\citep{S01} on a cubic lattice with lattice vectors ${\bf e}_i = (0,0,0)v_c$, $(\pm 1,0,0)v_c$, $(0, \pm 1,0)v_c$, $(0,0,\pm1)v_c$, $(\pm 1, \pm 1, \pm 1) v_c$. Physical variables are defined as moments of the distribution functions by 
\begin{equation}
\rho({\bf x},t) \equiv \sum_i g_i({\bf x},t), \,\, (\rho u_\alpha)({\bf x},t) \equiv \sum_i g_i({\bf x},t)  e_{i\alpha}.
\label{e:Definitions}
\end{equation}
The distribution functions evolve in time according to~\citep{BGK54}
\begin{equation}
D_i g_i \equiv \left( \partial_t + e_{i\alpha} \partial_\alpha \right) g_i = - \frac{1}{\tau} \left( g_i - g_i^\mathrm{eq} \right) + W_i,\label{e:EvolutionEquation}
\end{equation}where we have also defined the material derivative $D_i$ and a driving term $W_i$. By choosing appropriate moments for the equilibrium distribution $g_i^\mathrm{eq}$ and the driving term $W_i$ as
\begin{eqnarray}
&&\sum_i g_i^\mathrm{eq} = \rho; \qquad \sum_i g_i^\mathrm{eq} e_{i\alpha} = \rho u_\alpha;\nonumber\\
&&\sum_i g_i^\mathrm{eq} e_{i\alpha}e_{j\beta} = P_{\alpha\beta} + \rho u_\alpha u_\beta;  \label{e:Moments}\\
&&\sum_i W_i = 0; \qquad \sum_i W_i e_{i\alpha} = F_\alpha;\nonumber\\
&&\sum_i W_i e_{i\alpha} e_{j\beta} = u_\alpha F_\beta + F_\alpha u_\beta, \nonumber
\end{eqnarray}(\ref{eq:continuity}) and~(\ref{e:NavierStokes}) can be obtained from (\ref{e:EvolutionEquation}) via a Chapman-Enskog expansion similar to derivations in \citet{CD98}. The finite-difference scheme we use to solve (\ref{e:EvolutionEquation}) is discussed in \citet{ODKA11} (although thermal noise is not used here).  We emphasize that the results of the present work are by no means specific to this finite-difference algorithm, but it allows a stronger coupling between the fluid and the particle. In the present work, the particle is either held fixed or moved sinusoidally in which case its equation of motion can be solved analytically and used in the simulation.

\section{Polynomials $P_j$ in the solution to the uniform-density sphere}
\label{appendixb}
The polynomials in (\ref{eq:ourOmegaForSphere}) are reproduced in full below.  
\begin{align}
P_1 & = (Y^{12}+Z^{12})(81 + 162 Y + 162 Y^2 + 36 Y^3 + 4 Y^4) (3 + 3 Y + 2 \beta^2)\nonumber\\
& + 2 Y^4 Z^6 \bigl(243 + 729 Y + 972 Y^2 + 378 Y^3 - 324 Y^4 - 480 Y^5 - 312 Y^6 - 
 216 Y^7\nonumber\\
 & + 2 Y (27 + 54 Y + 48 Y^2 - 36 Y^4 - 8 Y^5) \beta^2\bigr)\nonumber\\
&  - Y^8 Z^2 \bigl(243 + 729 Y + 972 Y^2 + 1350 Y^3 + 1260 Y^4 + 888 Y^5 - 96 Y^6 + 
 168 Y^7\nonumber\\
& - 2 Y (81 + 162 Y + 144 Y^2 + 12 Y^3 - 24 Y^4 - 8 Y^5) \beta^2\bigr)\nonumber\\
& - Z^{10} \bigl(243 + 729 Y + 972 Y^2 + 702 Y^3 + 684 Y^4 + 1032 Y^5 + 912 Y^6 + 
   264 Y^7\nonumber\\
& - 2 Y (81 + 162 Y + 144 Y^2 - 12 Y^3 - 48 Y^4 - 8 Y^5) \beta^2\bigr)\nonumber\\
& - 3 Y^3 Z^8 \bigl(36 (3 + \beta^2) - Y (-321 - 651 Y - 732 Y^2 - 470 Y^3 + 8 Y^4 + 44 Y^5\nonumber\\
& + 2 (9 + 50 Y + 82 Y^2 + 28 Y^3 + 4 Y^4) \beta^2)\bigr)\nonumber\\
& + 3 Y^7 Z^4 \bigl(36 (3 + \beta^2) + Y (159 + 165 Y + 84 Y^2 + 74 Y^3 + 8 Y^4 + 44 Y^5\nonumber\\
& + 2 (81 + 130 Y + 98 Y^2 + 28 Y^3 + 4 Y^4) \beta^2)\bigr)
\end{align}
\begin{align}
P_2 & = Z^{10} \bigl(270 Y + 810 Y^2 + 1080 Y^3 + 648 Y^4 + 108 Y^5\nonumber\\
& + (81 + 180 Y + 198 Y^2 + 72 Y^3 + 8 Y^4) \beta^2\bigr)\nonumber\\
& + 4 Y^4 Z^6 \bigl(81 + 153 Y + 225 Y^2 + 198 Y^3 + 204 Y^4 + 54 Y^5\nonumber\\
& + (27 + 54 Y + 60 Y^2 + 30 Y^3 + 4 Y^4) \beta^2\bigr)\nonumber\\
& + Y^8 Z^2 \bigl(-324 - 234 Y + 234 Y^2 + 720 Y^3 + 264 Y^4 + 108 Y^5\nonumber\\ 
& + (27 + 84 Y + 90 Y^2 + 48 Y^3 + 8 Y^4) \beta^2\bigr)\nonumber\\
& + 2 Y^5 Z^4 \bigl(-243 - 729 Y - 972 Y^2 - 432 Y^3 + 204 Y^4 + 240 Y^5 - 36 Y^6\nonumber\\
&  - 2 Y (27 + 54 Y + 48 Y^2 + 18 Y^3 + 4 Y^4) \beta^2\bigr)\nonumber\\
& +Y^9 \bigl(243 + 729 Y + 972 Y^2 + 810 Y^3 + 492 Y^4 + 324 Y^5 - 12 Y^6\nonumber\\
& + (81 + 81 Y - 126 Y^3 - 36 Y^4 - 8 Y^5) \beta^2\bigr)\nonumber\\
& + Y Z^8 \bigl(243 + 729 Y + 972 Y^2 + 702 Y^3 + 252 Y^4 + 60 Y^5 - 60 Y^6\nonumber\\
& + (135 + 243 Y + 216 Y^2 - 18 Y^3 - 36 Y^4 - 8 Y^5) \beta^2\bigr)
\end{align}
\begin{align}
P_3 & = (Y^{12} + Z^{12}) (81 + 162 Y + 162 Y^2 + 36 Y^3 + 4 Y^4) (3 + 3 Y + 
   2 \beta^2)\nonumber\\
& + 2 Y^4 Z^6 \bigl(-243 - 729 Y - 972 Y^2 - 378 Y^3 + 324 Y^4 + 480 Y^5 + 312 Y^6 + 216 Y^7\nonumber\\
& + 2 Y (-27 - 54 Y - 48 Y^2 + 36 Y^4 + 8 Y^5) \beta^2\bigr)\nonumber\\
& + Z^{10} \bigl(243 + 729 Y + 972 Y^2 + 1350 Y^3 + 1260 Y^4 + 888 Y^5 - 96 Y^6 + 168 Y^7\nonumber\\
& + 2 Y (-81 - 162 Y - 144 Y^2 - 12 Y^3 + 24 Y^4 + 8 Y^5) \beta^2\bigr)\nonumber\\
& +Y^8 Z^2 \bigl(243 + 729 Y + 972 Y^2 + 702 Y^3 + 684 Y^4 + 1032 Y^5 + 912 Y^6 + 264 Y^7\nonumber\\
& + 2 Y (-81 - 162 Y - 144 Y^2 + 12 Y^3 + 48 Y^4 + 8 Y^5) \beta^2\bigr)\nonumber\\
& + 3 Y^7 Z^4 \bigl(-36 (3 + \beta^2) + Y(-321 - 651 Y - 732 Y^2 - 470 Y^3 + 8 Y^4 + 44 Y^5\nonumber\\ 
& + (18 + 100 Y + 164 Y^2 + 56 Y^3 + 8 Y^4) \beta^2)\bigr)\nonumber\\
& +3 Y^3 Z^8 \bigl(36 (3 + \beta^2) + Y (159 + 165 Y + 84 Y^2 + 74 Y^3 + 8 Y^4 + 44 Y^5\nonumber\\
& + (162 + 260 Y + 196 Y^2 + 56 Y^3 + 8 Y^4) \beta^2)\bigl)
\end{align}
\begin{align}
P_4 & = Y^{11} \bigl(270 Y + 810 Y^2 + 1080 Y^3 + 648 Y^4 + 108 Y^5\nonumber\\
& + (81 + 180 Y + 198 Y^2 + 72 Y^3 + 8 Y^4) \beta^2\bigr)\nonumber\\
& + 4 Y^7 Z^4 \bigl(81 + 153 Y + 225 Y^2 + 198 Y^3 + 204 Y^4 + 54 Y^5\nonumber\\
& + (27 + 54 Y + 60 Y^2 + 30 Y^3 + 4 Y^4) \beta^2\bigr)\nonumber\\
& + Y^3 Z^8 \bigl(-324 - 234 Y + 234 Y^2 + 720 Y^3 + 264 Y^4 + 108 Y^5\nonumber\\
& + (27 + 84 Y + 90 Y^2 + 48 Y^3 + 8 Y^4) \beta^2\bigr)\nonumber\\
& + 2 Y^4 Z^6 \bigl(243 + 729 Y + 972 Y^2 + 432 Y^3 - 204 Y^4 - 240 Y^5 + 36 Y^6\nonumber\\
& + (54 Y + 108 Y^2 + 96 Y^3 + 36 Y^4 + 8 Y^5) \beta^2\bigr)\nonumber\\
& + Y^8 Z^2 \bigl(-243 - 729 Y - 972 Y^2 - 702 Y^3 - 252 Y^4 - 60 Y^5 + 60 Y^6\nonumber\\
& + (-135 - 243 Y - 216 Y^2 + 18 Y^3 + 36 Y^4 + 8 Y^5)\beta^2\bigr)\nonumber\\
& + Z^{10} \bigl(-243 - 729 Y - 972 Y^2 - 810 Y^3 - 492 Y^4 - 324 Y^5 + 12 Y^6\nonumber\\
& + (-81 - 81 Y + 126 Y^3 + 36 Y^4 + 8 Y^5) \beta^2\bigr)
\end{align}
\begin{align}
P_5 & = (Y^{12} + Z^{12})(81 + 324 Y + 486 Y^2 + 360 Y^3 + 76 Y^4 + 8 Y^5)\nonumber\\
& + 2 Y^4 Z^6 \bigl(81 + 180 Y + 270 Y^2 + 216 Y^3 + 136 Y^4 - 72 Y^6 - 16 Y^7\nonumber\\
& + 2 (9 + 18 Y + 20 Y^2 + 16 Y^3 + 12 Y^4)\beta^2\bigr)\nonumber\\
& + Y^8 Z^2 \bigl(-81 - 72 Y + 270 Y^2 + 648 Y^3 + 536 Y^4 + 104 Y^5 - 40 Y^6 - 16 Y^7\nonumber\\ 
& + 2 (27 + 36 Y + 24 Y^2 - 12 Y^3 + 8 Y^4) \beta^2\bigr)\nonumber\\
& - Z^{10} \bigl(81 + 288 Y + 378 Y^2 + 216 Y^3 + 40 Y^4 + 104 Y^5 + 104 Y^6 + 16 Y^7\nonumber\\
& - 2 (27 + 72 Y + 96 Y^2 + 60 Y^3 + 16 Y^4) \beta^2\bigr)\nonumber\\
& + Y Z^8 \bigl(108 + 324 Y + 432 Y^2 + 63 Y^3 - 108 Y^4 + 58 Y^5 + 440 Y^6 + 164 Y^7 + 24 Y^8\nonumber\\
& + 4 Y \bigl(9 + 18 Y + 16 Y^2 - 4 Y^3 - 4 Y^4) \beta^2)\nonumber\\
& + Y^5 Z^4 \bigl(-108 - 324 Y - 432 Y^2 - 225 Y^3 + 324 Y^4 + 698 Y^5 + 568 Y^6 + 164 Y^7 + 24 Y^8\nonumber\\
& - 4 Y (9 + 18 Y + 16 Y^2 + 4 Y^3 + 4 Y^4 ) \beta^2\bigr)
\end{align}
\begin{align}
P_6 & = Y^8 Z^2 \bigl(108 + 324 Y + 432 Y^2 + 306 Y^3 + 144 Y^4 + 76 Y^5 + 44 Y^6 + 8 Y^7\nonumber\\
& - (45 + 99 Y + 108 Y^2 + 38 Y^3 + 12 Y^4) \beta^2\bigr)\nonumber\\
& +Y^3 Z^8 \bigl(81 + 288 Y + 378 Y^2 + 216 Y^3 - 32 Y^4 - 36 Y^5 - 8 Y^6\nonumber\\
& - (9 + 8 Y - 2 Y^2 - 8 Y^3) \beta^2\bigr)\nonumber\\
& + Z^{10} \bigl(90 Y^3 + 200 Y^4 + 220 Y^5 + 76 Y^6 + 8 Y^7 - 3 (9 + 27 Y + 36 Y^2 + 22 Y^3 + 4 Y^4) \beta^2 \bigr)\nonumber\\
& + Y^{11} \bigl(81 + 252 Y + 270 Y^2 + 72 Y^3 - 104 Y^4 - 36 Y^5 - 8 Y^6 - 3 (9 + 16 Y + 14 Y^2)\beta^2\bigr)\nonumber\\
& - 2 Y^7 Z^4 \bigl(81 + 162 Y + 216 Y^2 + 144 Y^3 + 100 Y^4 + 36 Y^5 + 8 Y^6\nonumber\\
& + (18 + 36 Y + 28 Y^2 - 4 Y^3) \beta^2\bigr)\nonumber\\
& - 4 Y^4 Z^6 \bigl(27 + 81 Y + 108 Y^2 + 45 Y^3 - 34 Y^4 - 58 Y^5 - 30 Y^6 - 4 Y^7\nonumber\\
& + Y (9 + 18 Y + 22 Y^2 + 6 Y^3) \beta^2\bigr)
\end{align}
\begin{align}
P_7 & = (Y^{12} + Z^{12}) (81 + 324 Y + 486 Y^2 + 360 Y^3 + 76 Y^4 + 8 Y^5)\nonumber\\
& - 2 Y^4 Z^6 \bigl(81 + 180 Y + 270 Y^2 + 216 Y^3 + 136 Y^4 - 72 Y^6 - 16 Y^7\nonumber\\
& + 2 (9 + 18 Y + 20 Y^2 + 16 Y^3 + 12 Y^4) \beta^2\bigr)\nonumber\\
& + Z^{10}\bigl(81 + 72 Y - 270 Y^2 - 648 Y^3 - 536 Y^4 - 104 Y^5 + 40 Y^6 + 16 Y^7\nonumber\\
& - 2 (27 + 36 Y + 24 Y^2 - 12 Y^3 + 8 Y^4) \beta^2\bigr)\nonumber\\
& +Y^8 Z^2 \bigl(81 + 288 Y + 378 Y^2 + 216 Y^3 + 40 Y^4 + 104 Y^5 + 104 Y^6 + 16 Y^7\nonumber\\
& - 2 (27 + 72 Y + 96 Y^2 + 60 Y^3 + 16 Y^4) \beta^2\bigr)\nonumber\\
& +Y^5 Z^4 \bigl(108 + 324 Y + 432 Y^2 + 63 Y^3 - 108 Y^4 + 58 Y^5 + 440 Y^6 + 164 Y^7 + 24 Y^8\nonumber\\
& + 2 Y (9 + 18 Y + 16 Y^2 - 4 Y^3 - 4 Y^4) \beta^2\bigr)\nonumber\\
& -Y Z^8 \bigl(108 + 324 Y + 432 Y^2 + 225 Y^3 - 324 Y^4 - 698 Y^5 - 568 Y^6 - 164 Y^7 - 24 Y^8\nonumber\\
& + 4 Y (9 + 18 Y + 16 Y^2 + 4 Y^3 + 4 Y^4) \beta^2\bigr)
\end{align}
\begin{align}
P_8 & = Y Z^8 \bigl(108 + 324 Y + 432 Y^2 + 306 Y^3 + 144 Y^4 + 76 Y^5 + 44 Y^6 + 8 Y^7\nonumber\\
& - (45 + 99 Y + 108 Y^2 + 38 Y^3 + 12 Y^4) \beta^2\bigr)\nonumber\\
& - Y^8 Z^2 \bigl(81 + 288 Y + 378 Y^2 + 216 Y^3 - 32 Y^4 - 36 Y^5 - 8 Y^6\nonumber\\
& - (9 + 8 Y - 2 Y^2 - 8 Y^3) \beta^2\bigr)\nonumber\\
& +Y^9 \bigl(90 Y^3 + 200 Y^4 + 220 Y^5 + 76 Y^6 + 8 Y^7\nonumber\\
& - (27 + 81 Y + 108 Y^2 + 66 Y^3 + 12 Y^4) \beta^2\bigr)\nonumber\\
& - Z^10 \bigl(81 + 252 Y + 270 Y^2 + 72 Y^3 - 104 Y^4 - 36 Y^5 - 8 Y^6 - 3 (9 + 16 Y + 14 Y^2) \beta^2\bigr)\nonumber\\
& + 2 Y^4 Z^6 \bigl(81 + 162 Y + 216 Y^2 + 144 Y^3 + 100 Y^4 + 36 Y^5 + 8 Y^6\nonumber\\
& + (18 + 36 Y + 28 Y^2 - 4 Y^3) \beta^2\bigr)\nonumber\\
& - 4 Y^5 Z^4 \bigl(27 + 81 Y + 108 Y^2 + 45 Y^3 - 34 Y^4 - 58 Y^5 - 30 Y^6 - 4 Y^7\nonumber\\
& + (9 Y + 18 Y^2 + 22 Y^3 + 6 Y^4) \beta^2\bigr)
\end{align}
\begin{align}
P_9 & = (Z^{10} - Y^{10}) (9 + 18 Y + 18 Y^2 + 36 Y^3 + 36 Y^4 + 12\beta^2 + 12 Y\beta^2 + 4\beta^4)\nonumber\\
& - Z^8 \bigl(9 + 18 Y + 9 Y^2 + 18 Y^3 + 54 Y^4 + 36 Y^5 + 36 Y^6\nonumber\\
& - 12 Y (1 + 3 Y + 3 Y^2)\beta^2 + 4 Y^2 \beta^4\bigr)\nonumber\\
& + Y^6 Z^2 \bigl(9 + 18 Y + 9 Y^2 + 18 Y^3 - 90 Y^4 - 108 Y^5 + 36 Y^6\nonumber\\
& - 12 Y(1 + 3 Y - Y^2) \beta^2 + 4 Y^2 \beta^4\bigr)\nonumber\\
& +Y^4 Z^4 \bigl(9 + 18 Y + 18 Y^2 - 36 Y^3 - 216 Y^4 - 72 Y^5 - 72 Y^6\nonumber\\
& - 12 Y (3 + 2 Y - 2 Y^2) \beta^2 - 8 Y^2 \beta^4\bigr)\nonumber\\
& -Y^2 Z^6 \bigl(9 + 18 Y + 18 Y^2 + 108 Y^3 + 216 Y^4 + 72 Y^5 - 72 Y^6\nonumber\\
& + 12 Y (1 - 2 Y - 2 Y^2)\beta^2 - 8 Y^2 \beta^4\bigr)
\end{align}
\begin{align}
P_{10} & = 2 Z^8 (9 Y + 18 Y^2 + 18 Y^3 + 18 Y^4 + 2 \beta^4)\nonumber\\
& - 3 Y^5 Z^2 \bigl(3 + 6 Y + 24 Y^2 + 24 Y^3 - 12 Y^4 + 2 (1 - 4 Y - 2 Y^2) \beta^2\bigr)\nonumber\\
& + 3 Y Z^6 \bigr(3 + 6 Y + 12 Y^2 + 24 Y^3 + 12 Y^4 + 2 (3 + 4 Y + 2 Y^2) \beta^2\bigr)\nonumber\\
& - Y^7 \bigl(9 + 18 Y + 18 Y^2 + 36 Y^3 - 36 Y^5 + 6 (1 - 2 Y^2) \beta^2 - 4 Y \beta^4\bigr)\nonumber\\
& +Y^3 Z^4 \bigl(9 + 18 Y - 36 Y^2 + 36 Y^4 + 72 Y^5 + 6 (3 + 8 Y + 2 Y^2) \beta^2 + 8 Y \beta^4\bigr)
\end{align}
\begin{align}
P_{11} & = (Y^{10} + Z^{10}) (9 + 18 Y + 18 Y^2 + 36 Y^3 + 36 Y^4 + 12 (1 + Y) \beta^2 + 4 \beta^4)\nonumber\\
& + Y^6 Z^2 \bigl(9 + 18 Y + 9 Y^2 + 18 Y^3 + 54 Y^4 + 36 Y^5 + 36 Y^6\nonumber\\
& - 12 Y (1 + 3 Y + 3 Y^2) \beta^2 + 4 Y^2 \beta^4\bigr)\nonumber\\
& + Z^8 \bigl(9 + 18 Y + 9 Y^2 + 18 Y^3 - 90 Y^4 - 108 Y^5 + 36 Y^6\nonumber\\
& - 12 Y (1 + 3 Y - Y^2) \beta^2 + 4 Y^2 \beta^4\bigr)\nonumber\\
& - Y^2 Z^6 \bigl(9 + 18 Y + 18 Y^2 - 36 Y^3 - 216 Y^4 - 72 Y^5 - 72 Y^6\nonumber\\
& - 4 Y (9 + 6 Y - 6 Y) \beta^2 - 8 Y^2 \beta^4\bigr)\nonumber\\
& -Y^4 Z^4 \bigl(9 + 18 Y + 18 Y^2 + 108 Y^3 + 216 Y^4 + 72 Y^5 - 72 Y^6\nonumber\\
& + 12 Y (1 - 2 Y - 2 Y^2) \beta^2 - 8 Y^2 \beta^4\bigr)
\end{align}
\begin{align}
P_{12} & = 2 Y^9 (9 Y + 18 Y^2 + 18 Y^3 + 18 Y^4 + 2 \beta^4)\nonumber\\
& + 3 Y^2 Z^6 \bigl(3 + 6 Y + 24 Y^2 + 24 Y^3 - 12 Y^4 + 2 (1 - 4 Y - 2 Y^2) \beta^2\bigr)\nonumber\\
& - 3 Y^6 Z^2 \bigl(3 + 6 Y + 12 Y^2 + 24 Y^3 + 12 Y^4 + 2 (3 + 4 Y + 2 Y^2) \beta^2\bigr)\nonumber\\
& - Z^8 \bigl(9 + 18 Y + 18 Y^2 + 36 Y^3 - 36 Y^5 + 6 (1 - 2 Y^2) \beta^2 - 4 Y \beta^4\bigr)\nonumber\\
& + Y^4 Z^4 \bigl(9 + 18 Y - 36 Y^2 + 36 Y^4 + 72 Y^5 + 6 (3 + 8 Y + 2 Y^2) \beta^2 + 8 Y \beta^4\bigr)
\end{align}



\begin{thebibliography}{47}

\bibitem[Abade \etal(2010a)]{ACE-JNW10}
{\sc Abade, G.C., Cichocki, B., Ekiel-Jezewska, M.~L., N\"agele, G. \& Wajnryb, E.} 2010a Dynamics of permeable particles in concentrated suspensions, {\em Phys. Rev. E.\/} {\bf 81}, 020404(R).

\bibitem[Abade \etal(2010b)]{ACE-JNW10b}
{\sc Abade, G.C., Cichocki, B., Ekiel-Jezewska, M.~L., N\"agele, G. \& Wajnryb, E.} 2010b Short-time dynamics of permeable particles in concentrated suspensions, {\em J. Chem. Phys.\/} {\bf 132}, 014503.

\bibitem[Ahlrichs \& D\"{u}nweg(1998)]{AD98}
{\sc Ahlrichs P. \& D\"{u}nweg, B.} 1998 Lattice Boltzmann simulation of polymer-solvent systems {\em International Journal of Modern Physics C} {\bf 9}, 1429; 1999 Simulation of a single polymer chain in solution by combining lattice Boltzmann and molecular dynamics {\em J. Chem. Phys} {\bf 111}, 8225.

\bibitem[Amro \& Al-Homadhi(2006)]{AA-H06}
{\sc Amro, M.M. \& Al-Homadhi, E.S.} 2006 Enhanced oil recovery using sound-wave stimulation {\em Final Research Report No. 53/426}, King Saud University.

\bibitem[Babadagli(2003)]{Baba03}
{\sc Babadagli, T.} 2003 Selection of proper enhanced oil recovery fluid for efficient matrix recovery in fractured oil reservoirs {\em Colloids and Surfaces A\/} {\bf 223}, 157--175.

\bibitem[Batchelor(1967)]{Batchelor67}
{\sc Batchelor, G. K.} 1967 {\em An Introduction to Fluid Mechanics\/}, Cambridge University Press UK.

\bibitem[Beavers \& Joseph(1967)]{BJ67}
{\sc Beavers, G. S. \& Joseph, D. D.} 1967 Boundary conditions at a naturally permeable wall, {\em J. Fluid. Mech.\/} {\bf 30}, 197--207.

\bibitem[Bergendahl \& Grasso(2000)]{BG00}
{\sc Bergendahl J. \& Grasso D.} 2000 Prediction of Colloid Detachment in a Model Porous Media: Hydrodynamics {\em Chem. Eng. Sci.\/} {\bf 55}, 1523--1532.

\bibitem[Bhatt \& Sacheti(1994)]{BS93}
{\sc Bhatt, B. S. \& Sacheti, N.C.} 1994 Flow past a porous spherical shell using the Brinkman model {\em J. Phys. D: Appl. Phys.\/} {\bf 27}, 37--41.

\bibitem[Bhatnagar \etal(1954)]{BGK54}
{\sc Bhatnagar, P. L., Gross, E. P. \& M. Krook, M.} 1954 A Model for Collision Processes in Gases. I. Small Amplitude Processes in Charged and Neutral One-Component Systems {\em Phys. Rev.} {\bf 94}, 511.

\bibitem[Brinkman(1947a)]{B47_1}
{\sc Brinkman, H. C.} 1947a A calculation of the viscous force exerted by by a flowing fluid on a dense swarm of particles, {\em Appl. Sci. Res.\/} {\bf A1}, 27--34.

\bibitem[Brinkman(1947b)]{B47_2}
{\sc Brinkman, H. C.} 1947b On the permeability of media consisting of closely packed porous particles, {\em Appl. Sci. Res.\/} {\bf A1}, 81--86.

\bibitem[Burden \& Faires(2005)]{BF05}
{\sc Burden, R. L. \& Faires, J. D.} 2005 {\em Numerical Analysis, 8th ed.}, Thomson, USA.

\bibitem[Chen \& Doolen(1998)]{CD98}
{\sc Chen, S. \& Doolen, G.D.} 1998 Lattice Boltzmann method for fluid flows {\em Annual Review of Fluid Mechanics\/} {\bf 30}, 329.

\bibitem[Cichocki \& Felderhof(2009)]{CF09}
{\sc Cichocki, B. \& Felderhof, B. U.} 2009 Hydrodynamic friction coefficients of coated spherical particles {\em J. Chem. Phys.\/} {\bf 130}, 164712.

\bibitem[Debye \& Bueche(1948)]{DB48}
{\sc Debye, P. \& Bueche, A. M.} 1948 Intrinsic Viscosity, Diffusion, and Sedimentation Rate of Polymers in Solution {\em J. Chem. Phys.\/} {\bf 16}, 573.

\bibitem[Del Bonis-O'Donnell \etal(2009)]{Stein09}
{\sc Del Bonis-O'Donnell, J.~T., Reisner, W. \& Stein, D.} 2009 Pressure-driven DNA transport across an artificial nanotopography, {\em New J. Phys.\/} {\bf 11}, 075032.

\bibitem[Deutch \& Felderhof(1975)]{FD75b}
{\sc Deutch, J. M. \& Felderhof, B. U.} 1975 Frictional properties of dilute polymer solutions. II. The effect of preaveraging {\em J. Chem. Phys.\/} {\bf 62}, 2398.

\bibitem[Edwards(1997)]{RLanger97}
{\sc Edwards,D.~A.} \etal\/ 1997 Large Porous Particles for Pulmonary Drug Delivery, {\em Science\/} {\bf 276}, 1868--1872.

\bibitem[Felderhof(1975)]{F75}
{\sc Felderhof, B. U.} 1975 Frictional properties of dilute polymer solutions: III. Translational-friction coefficient {\em Physica\/} A {\bf 80}, 63--75.

\bibitem[Felderhof \& Deutch(1975)]{FD75a}
{\sc Felderhof, B.U. \& Deutch, J.M.} 1975 Frictional properties of dilute polymer solutions I. Rotational friction coefficient {\em J. Chem. Phys.\/} {\bf 62}, 2391.

\bibitem[Fu \etal(2007)]{Fu07}
{\sc Fu, J.P., Schoch, R.~B., Stevens, A.~L., Tannenbaum, S.~R. \& Han, J.~Y.} 2007 A patterned anisotropic nanofluidic sieving structure for continuous-flow separation of DNA and proteins,  {\em Nat. Nanotechnol.\/} {\bf 2}, 121.
 
\bibitem[Goldman \etal(1967)]{Goldman67b}
{\sc Goldman, A., Cox R. \& Brenner, H.} 1967 Slow viscous motion of a sphere parallel to a plane wall - II Couette flow, {\em Chem. Eng. Sci.} {\bf  22}, 653.

\bibitem[Graebel(2007)]{G07}
{\sc Graebel, W. P.} 2007 {\em Advanced Fluid Mechanics\/}, Academic Press USA.

\bibitem[Higuera \etal(1989)]{HSB89}
{\sc Higuera, F., Succi, S. \& Benzi, R.} 1989 Lattice gas dynamics with enhanced collisions {\em Europhys. Lett.} {\bf 9} 345.

\bibitem[J\"ager \& Mikeli\'c(2000)]{JM00}
{\sc J\"ager W. \& Mikeli\'c, A.} 2000 On the interface boundary condition of Beavers, Joseph, and Saffman, {\em SIAM Soc. Ind. Appl. Math. J. Appl. Math.\/} {\bf 60}, 1111.

\bibitem[Jimenez(1989)]{HJ89}
{\sc Jimenez, F. H. J.} 1989 Boltzmann approach to lattice gas simulations {\it Europhys. Lett.} {\bf 9} 663.

\bibitem[Kell(1970)]{Kell70}
{\sc Kell, G. S.} 1970 Isothermal compressibility of liquid water at 1 atm, {\em J. Chem. Eng. Data} {\bf 15}, 119--122.

\bibitem[Kirkwood \& Riseman(1948)]{KR48}
{\sc Kirkwood, J. G. \& J. Riseman, J.} 1948 The intrinsic viscosities and diffusion constants of flexible macromolecules in solution{\em J. Chem. Phys.} {\bf 16}, 565.

\bibitem[Lamb(1932)]{L32}
{\sc  Lamb, H.} 1932 {\em Hydrodynamics 6th ed.}, MacMillan, New York.

\bibitem[Landau \& Lifschitz(1987)]{LandauLifschitz87}
{\sc Landau, L.D. \& Lifschitz, E.M.} 1987 {\em Fluid Mechanics 2nd ed.\/}, Pergamon Press.

\bibitem[Liron \& Mochon(1976)]{LM76}
{\sc Liron N. \& Mochon, S.} 1976 Stokes flow for a stokeslet between two parallel flat plates {\em J. Eng. Math.} {\bf 10}, 287.

\bibitem[Looker \& Carnie(2004)]{LC04}
{\sc Looker, J. R. \& Carnie, S. L.} 2004 The hydrodynamics of an oscillating porous sphere {\em Phys. Fluids\/} {\bf 16}, 62--72.

\bibitem[Lundgren \etal(1972)]{Lundgren72}
{\sc Lundgren, T. S.} 1972 Slow flow through stationary random beds and suspensions of spheres {\it J. Fluid. Mech.} {\bf 51} 273--299.

\bibitem[Neale \etal(1973)]{NEN73}
{\sc Neale, G., Epstein, N. \& Nader, W.} 1973 Creeping flow relative to permeable spheres {\em Chem. Eng. Sci.} {\bf 28}, 1865.

\bibitem[Ollila \etal(2011a)]{ODKA11}
{\sc Ollila, S. T. T., Denniston, C., Karttunen, M. \& Ala-Nissila, T.} 2011a Fluctuating lattice-Boltzmann model for complex fluids {\it J. Chem. Phys.\/} {\bf 134}, 064902.

\bibitem[Ollila \etal(2011b)]{OSD11}
{\sc Ollila, S. T. T., Smith,C. J., Ala-Nissila, T. \& Denniston, C.} 2011b {\em unpublished}.

\bibitem[Peskin(2002)]{P02}
{\sc Peskin, C. S.} 2002 The immersed boundary method, {\it Acta Numerica} {\bf 11}, 479.

\bibitem[Kang \etal(2002)]{KZC02}
{\sc Kang, Q., Zhang, D. \& Chen S.} 2002 Unified lattice Boltzmann method for flow in multiscale porous media {\em Phys. Rev. E.\/} {\bf 66}, 056307.

\bibitem[Ramaswamy \etal(2004)]{RGGAKKR04}
{\sc Ramaswamy, S., Gupta, M., Goela, A., Aaltosalmi, U., Kataja, M., Koponen, A., \& Ramarao, B.V.} 2004 Efficient simulation of flow in and transport in porous media {\em Colloids and Surfaces A\/} {\bf 241}, 323--333.

\bibitem[Smith \& Denniston(2007)]{SD07}
{\sc Smith, C. J. \& Denniston, C.} 2007 Elastic response of a nematic liquid crystal to an immersed nanowire,  {\em J. Appl. Phys.\/} {\bf 101}, 014305.

\bibitem[Sparreboom \etal(2010)]{SvdBE10}
{\sc Sparreboom, W., van der Berg, A. \& Eijkel, J.~C.~T.}, 2010 Transport in nanofluidic systems: a review of theory and applications, {\em New J. Phys.\/} {\bf 12}, 015004.

\bibitem[Stokes(1880)]{S80}
{\sc Stokes, G.} 1880 {\em Mathematical and Physical Papers Volume I} located at www.archive.org/details/mathphyspapers01stokrich, Cambridge University Press.

\bibitem[Stokes(1901)]{S51}
{\sc Stokes, G. G.} 1901 {\em Mathematical and Physical Papers Volume III}, Cambridge University Press.

\bibitem[Succi(2001)]{S01}
{\sc Succi, S.} 2001 {\it The Lattice Boltzmann Equation for Fluid Dynamics and Beyond\/}, Oxford University Press, New York.

\bibitem[Sukop \& Thorne(2006)]{ST05}
{\sc Sukop, M. C. \& Thorne, D. T. J.} 2006 {\em Lattice Boltzmann Modeling: An Introduction for Geoscientists and Engineers\/}, Springer-Verlag Berlin Heidelberg.

\bibitem[Sutherland \& Tan(1970)]{ST70}
{\sc Sutherland, D. N. \& Tan, C. T.} 1970 Sedimentation of a porous sphere {\em Chem. Eng. Sci.} {\bf 25}, 1948--1950.

\bibitem[Swift \etal(1995)]{SOY95}
{\sc Swift, M. R., Osborn, W. R. \& J. M. Yeomans, J. M.} 1995 {\em Phys. Rev. Lett.} 1995 {\bf 75}, 830.

\bibitem[Vainshtein \& Shapiro(2009)]{VS09}S. L.
{\sc Vainshtein, P. \& Shapiro, M.} 2009 {\em Journal of Colloid and Interface Science\/} {\bf 330}, 149--155.

\bibitem[Vainshtein \etal(2002)]{VSG02}
{\sc Vainshtein, P., Shapiro, M. \& Gutfinger, C.} 2002 Creeping flow past and within a permeable spheroid, {\em Int. J. Multiphase Flow} {\bf 28}, 1945--1963.

\bibitem[Womersley(1955)]{W55}
{\sc Womersley J. R.} 1955 Method for the calculation of velocity, rate of flow and viscous drag in arteries when the pressure gradient is known, {\em J. Physiol.} {\bf 127}, 553--563.


\end{thebibliography}
\end{document}